\documentclass[a4paper,11pt,pdftex]{article}
\pdfoutput=1

\usepackage{slashed}
\usepackage{braket}
\usepackage{jheppub}
\usepackage{graphicx}
\usepackage{url}
\usepackage{amssymb,amsmath}
\usepackage{float}
\usepackage{wrapfig}
\usepackage{bm}
\usepackage{CJKutf8}
\usepackage{color}
\usepackage{comment}
\usepackage{xcolor}
\usepackage[normalem]{ulem}
\usepackage{hyperref}
\usepackage{mathrsfs}
\usepackage[version=4]{mhchem}
\usepackage{booktabs}
\usepackage[noabbrev]{cleveref}
\hypersetup{
colorlinks=true,
citecolor=blue,
citebordercolor=red,
linkcolor=blue,
urlcolor=blue
}

\catcode`\@=11
\catcode`\@=12
\newcommand{\lsim}{\lesssim}

\newcommand{\p}{\partial}

\newcommand{\bp}{\begin{pmatrix}}
\newcommand{\ep}{\end{pmatrix}}
\newcommand{\nn}{\nonumber\\}

\newcommand{\df}{\text{d}}

\newcommand{\bs}[1]{\boldsymbol}

\newcommand{\Tr}{{\rm Tr}\,}

\newcommand{\pmat}[1]{\begin{pmatrix}#1\end{pmatrix}}

\newcommand{\n}{\nonumber}

\newcommand{\mr}[1]{\mathrm{#1}}

\newcommand{\be}{\begin{equation}}
\newcommand{\ee}{\end{equation}}

\newcommand{\ba}{\begin{array}} 
\newcommand{\ea}{\end{array}}

\graphicspath{{./figs/}}
\newbox{\ORCIDicon}
\sbox{\ORCIDicon}{\large \includegraphics[width=0.8em]{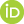}}

\makeatletter
\gdef\@fpheader{\phantom{prepared for submission to JHEP}}
\makeatother

\makeatletter
\@addtoreset{equation}{section}
\makeatother

\begin{document}
\begin{flushright} 
KEK-TH-2592, DESY-24-003, RESCEU-2/24, YGHP-24-01
\end{flushright} 
\title{
Neutrino zeromodes on electroweak strings in light of topological insulators
}
\author[a,b,c]{Minoru Eto,\,\href{https://orcid.org/0000-0002-2554-1888}{\usebox{\ORCIDicon}}}
\emailAdd{meto@sci.kj.yamagata-u.ac.jp}
\affiliation[a]{Department of Physics, Yamagata University, Kojirakawa-machi 1-4-12, Yamagata, Yamagata 990-8560, Japan}
\affiliation[b]{Research and Education Center for Natural Sciences, Keio University, 4-1-1 Hiyoshi, Yokohama, Kanagawa 223-8521, Japan}
\affiliation[c]{
International Institute for Sustainability with Knotted Chiral Meta Matter(SKCM$^2$), Hiroshima University, 1-3-2 Kagamiyama, Higashi-Hiroshima, Hiroshima 739-8511, Japan
}
\author[d,e,b]{Yu Hamada,\,\href{https://orcid.org/0000-0002-0227-5919}{\usebox{\ORCIDicon}}}
\emailAdd{yu.hamada@desy.de}
\affiliation[d]{Deutsches Elektronen-Synchrotron DESY, Notkestr. 85, 22607 Hamburg, Germany}
\affiliation[e]{KEK Theory Center, Tsukuba 305-0801, Japan}

\author[f]{Ryusuke Jinno,\,\href{https://orcid.org/0000-0003-0949-6623}{\usebox{\ORCIDicon}}}
\emailAdd{ryusuke.jinno@resceu.s.u-tokyo.ac.jp}
\affiliation[f]{Research Center for the Early Universe (RESCEU), University of Tokyo, Hongo 7-3-1, Bunkyo-ku, Tokyo 113-003, Japan
}

\author[g,b,c]{Muneto Nitta,\,\href{https://orcid.org/0000-0002-3851-9305}{\usebox{\ORCIDicon}}}
\emailAdd{nitta@phys-h.keio.ac.jp}
\affiliation[g]{Department of Physics, Keio University, 4-1-1 Hiyoshi, Kanagawa 223-8521, Japan}
\author[h]{and Masatoshi Yamada\,\href{https://orcid.org/0000-0002-1013-8631}{\usebox{\ORCIDicon}}}
\emailAdd{yamada@jlu.edu.cn}
\affiliation[h]{Center for Theoretical Physics and College of Physics, Jilin University, Changchun 130012, China}

\abstract{
We examine neutrino zeromode solutions on the electroweak $Z$-string and their effect on the stability of the string in the standard model and its extensions.
We propose using topological invariants constructed from the momentum (and real) space topology of Green’s functions, often used for investigating edge modes in condensed matter physics.
We analyze the standard model and then examine type-I and type-II extensions of the neutrino sector as well as their hybrid.
Based on this analysis, we also comment on proposals in the literature to stabilize the $Z$-string.
}


\maketitle 

\section{Introduction}

Spontaneous symmetry breaking is a widely observed phenomenon in statistical physics and quantum field theory. Its notion is crucial for understanding critical phenomena such as superconductivity~\cite{Bardeen:1957kj} in condensed matter physics and dynamical mass generation~\cite{Nambu:1961tp,Nambu:1961fr,Higgs:1964pj,Englert:1964et,Guralnik:1964eu} in high energy physics. 
A basic point in spontaneous symmetry breaking is that the trivial vacuum (or symmetric phase) associated to a symmetry in a system becomes an unstable (tachyonic) or metastable state which indicates existence of a lower energy state (broken phase) not obeying the symmetry. 
Such a simple phase-transition picture in spontaneous symmetry breaking can be described by the Ginzburg-Landau theory~\cite{Ginzburg:1950sr} with the free energy as a function of an order parameter associated to a symmetry. 
In the broken phase with non-vanishing order parameter value, 
the equation of motion (EOM) could entail non-trivial solutions depending on spacial coordinates in addition to the vacuum solutions. 
One of well-known solutions is an Abrikosov-Nielsen–Olesen (ANO) vortex/string solution~\cite{Abrikosov:1956sx,Nielsen:1973cs} in a $U(1)$ gauge theory coupled with one complex scalar field, called the Abelian-Higgs model.\footnote{
The ANO vortex/string solution has been mainly studied for the scalar potential in polynomials of the scalar field $\#|\Phi|^2+\#|\Phi|^4$. Recently, we have studied the ANO vortex/string solution in the Coleman-Weinberg type potential $|\Phi|^4\log (|\Phi|^2/v_\Phi^2)$ in Ref.~\cite{Eto:2022hyt} and have shown non-trivial interactions between two vortices/strings depending on mass ratio between the scalar and gauge boson masses.  
}
In addition, there are also other objects such as monopoles, domain walls, and Skyrmions.
These solutions arise due to the existence of the non-trivial homotopy group characterizing maps between the order-parameter space (vacuum manifold) and the real space.
Such topologically non-trivial solutions are called topological defects or solitons. 
In the context of particle cosmology, phase transitions result in creation of topological defects through the Kibble-Zurek mechanism~\cite{Kibble:1976sj,Zurek:1985qw}. 
These objects may give great impacts on the thermal history of the universe.

If topological defects couple with fermionic fields, 
they may lead to non-trivial phenomena.
In fact, it is well known that fermionic modes sometimes are trapped inside the defects.
For example, fermionic zeromodes localized on vortices in topological superconductors~\cite{PhysRevB.44.9667,Volovik:1999eh,Read:1999fn,Ivanov:2000mjr,Fukui:2009mh,Volovik:2003fe},
and Majorana fermion zeromodes on superfluid vortices~\cite{Sedrakian:2018ydt} in neutron stars~\cite{Masaki:2019rsz,Masaki:2021hmk,Masaki:2023rtn}
and on non-Abelian vortices~\cite{Nishida:2010wr,Yasui:2010yw,Fujiwara:2011za,Eto:2013hoa,Chatterjee:2016tml,Zubkov:2016llc} in quantum chromodynamics (QCD) in high density region (quark matter)~\cite{Balachandran:2005ev,Nakano:2007dr,Eto:2009bh,Eto:2009tr,Eto:2013hoa}.
Whether the zeromodes exist on the defects or not reflects different phases of matter
that cannot be captured by the conventional Ginzburg-Landau theories. 
These phases are called topological phases and are classified by the homotopy groups of the \textit{momentum space} of the fermions. 
The relation between the zeromodes and the momentum space topology has been widely studied utilizing Green's function ~\cite{Volovik:2003fe,gurarie2011single,Volovik:2016mre,Zubkov:2016llc,PhysRevB.84.125132}
and (one-particle) Hamiltonian~\cite{Schnyder:2008tya,Kitaev:2009mg,schnyder2009classification,Teo:2010zb,Ryu:2010zza}.
Thus, in such systems, it is important to consider the topology of both of real space and momentum space.

In contrast to these condensed matter, nuclear and quark matter systems, 
all of which are finite density matter and are therefore {\it non-relativistic},   
vortex strings in {\it relativistic} field theories 
also contain fermion zeromodes~\cite{Nohl:1975jg,deVega:1976rt,Jackiw:1981ee,Weinberg:1981eu,Callan:1984sa}. 
Their qualitative properties are understood by the index theorem counting the number of zeromodes~\cite{Weinberg:1981eu,Semenoff:1987ki,Ganoulis:1987np}
and the anomaly inflow \cite{Callan:1984sa,Bagherian:2023jxy}.
In addition, Bogomol'nyi-Prasad-Sommerfield (BPS) vortices in supersymmetric field theories 
are always associated with fermion zeromodes 
as superpartners of translational (bosonic) zeromodes~\cite{Eto:2006pg,Shifman:2007ce,Shifman:2009zz}.
In the context of cosmology, fermion zeromodes on cosmic strings are also studied~\cite{Hindmarsh:1991ax} 
in application for instance to superconducting cosmic strings~\cite{Witten:1984eb}.

The fermionic zeromodes might play important roles 
even in the electroweak theory of the Standard Model (SM) in particle physics.
In fact, the electroweak theory admits a string-like solution of the EOMs
called the electroweak $Z$-string~\cite{Nambu:1977ag,Vachaspati:1992fi,Achucarro:1999it}, consisting of the $Z$ boson and Higgs fields.
This is a non-topological soliton and its stability is not topologically ensured.
Indeed, its stability is highly limited to the small parameter region~\cite{James:1992zp,James:1992wb,Goodband:1995he}.\footnote{Several mechanisms to stabilize the electroweak string or to enlarge its stable region have been suggested: (i) effective potential at finite-temperature~\cite{Holman:1992rv}; (ii) strong magnetic field~\cite{Garriga:1995fv}; (iii) thermal photon plasma~\cite{Nagasawa:2002at}; (iv) axion in core of the $Z$-string~\cite{Masperi:1993fw}; (v) dark scalar condensation~\cite{Forgacs:2019tbn}; 
(vi) splitting the $Z$-string into two fractional $Z$-strings in the two-Higgs doublet model~\cite{Eto:2021dca}.
}
See also Fig.~\ref{fig:stable region} in Section~\ref{sec: Stability of $Z$-string} for the stability region of the electroweak string. 
The electroweak strings, however, have attracted attention because they are proposed to be useful for baryogenesis \cite{Brandenberger:1992ys,Barriola:1994ez,Vachaspati:1994ng} and generation of cosmological magnetic fields \cite{Vachaspati:2001nb,Poltis:2010yu}.
Because the SM fermions inevitably couple with the $Z$ boson and Higgs fields,
the $Z$-string can trap fermion zeromodes inside them, 
and their effects on the stability of the $Z$-strings has been studied so far~\cite{Vachaspati:1992mk,Earnshaw:1994jj,Moreno:1994bk,Garriga:1994wb,Naculich:1995cb,Kono:1995xp,Groves:1999ks,Volovik:2015llj}. 
In particular, the existence of the zeromodes was originally proposed to improve the stability~\cite{Vachaspati:1992mk}
while it was argued~\cite{Naculich:1995cb,Kono:1995xp,Liu:1995at} that it rather destabilizes 
because the zeromodes appear as pairs of right-mover and left-mover modes along the $Z$-string 
and easily form massive modes to lower the energy of the configuration.

The SM introduces only the left-handed neutrinos to be massless. 
Turning the eye to the phenomenological aspects of high energy physics and cosmology,
neutrino oscillation phenomena have been observed and require at lease two of the three neutrinos to be massive~\cite{ParticleDataGroup:2022pth}.
Although the neutrino-less double beta decay is expected to provide a smoking gun for the distinction between Dirac and Majorana masses for the neutrinos, 
it is not yet observed.
Thus there is still a lot of possibilities for extensions of the neutrino sector. 
Such extensions could result in a finite number of zeromodes (or here equivalently topological number) to stabilize the electroweak string. 
The simplest extension would be the inclusion of the right-handed neutrinos with Majorana masses underlying the seesaw mechanism. 
It is found in Ref.~\cite{Starkman:2000bq} that in the simplest extension, the Dirac equations for the left- and right-handed neutrino fields have a (normalizable) zeromode solution. Together with a zeromode of the electron field, this indicates that the electroweak string is not stabilized. The zeromode counting is performed in another extension of the SM~\cite{Starkman:2001tc}.

In this work, we revisit the neutrino zeromodes on the electroweak-string background in several extensions of the neutrino sector. 
In particular, we aim to understand the non-trivial topology of the electroweak string arising due to the existence of fermion zeromodes in terms of the topological invariant in momentum space of Green's functions~\cite{Volovik:2003fe,Volovik:2016mre,Zubkov:2016llc,PhysRevB.84.125132}.\footnote{More precisely,
we utilize the topological invariant defined in momentum and real spaces since Green's function contains both the momenta and real space coordinates in presence of vortices.
Nevertheless, to emphasize the dependence on the momenta, we use the word ``momentum space topology'' in this paper.} 
This way allows us to directly count the number of zeromodes from the topology of momentum space without looking for explicit zeromode solutions. 
Therefore, we do not have to worry about missing the number of zeromodes. 
Such a topological invariant has been mainly used in non-relativistic systems stated above. 
This method has several advantages compared to the argument based on the index theorem~\cite{Weinberg:1981eu,Semenoff:1987ki,Ganoulis:1987np}, which is more familiar to high energy physicists.
Firstly, the arguments in those papers do not cover cases with Majorana masses
while our method relies only on Green's functions and is general enough to cover Majorana cases.
Secondly, our method directly calculates the topological invariant defined on real and momentum spaces,
which allows us to deform the setup continuously as long as it does not change the topology.
We intend to employ this also in particle physics and then demonstrate results obtained in several possible extensions of the neutrino sector including the early studies~\cite{Starkman:2000bq,Starkman:2001tc} in order to show that this method successfully counts the number of zeromodes.

This paper is organized as follows: In Section~\ref{sec:zeromode_review}, after summarizing model setups together with clarification of our notation, we briefly review the electroweak string (or $Z$-string) in the SM and its instability under small perturbations. 
In Section~\ref{sec:momentum_space_topology_1d}, we introduce the topological invariant that counts the number of fermion zeromodes appearing on the $Z$-string solutions.
We apply this method to several models as extended neutrino sectors.
By comparing these with explicit results obtained by solving the Dirac equations,
we show that our method gives the correct values of the number of the zeromodes.
Section~\ref{sec:discussion_conclusions} is devoted to discussion and conclusions.
In Appendix~\ref{sec:momentum_space_topology_3d}, we introduce the topological number in terms of the momentum-space topology~\cite{Groves:1999ks,Volovik:2016mre,Zubkov:2016llc} and evaluate it in several models for the neutrino sectors.

\section{Review on electroweak string in SM}
\label{sec:zeromode_review}
In this section, we briefly review the electroweak string solution (or the $Z$-string solution) in the SM sector.

\subsection{Model setup and electroweak string solutions}

We consider a string solution in the SM. It is known as the electroweak string or the $Z$-string, which has been first argued by Nambu in Ref.~\cite{Nambu:1977ag}.

We start with the setup of the model. For our purpose, we focus on the first generation of the leptons (i.e. $e^-$ and $\nu_{eL}$), $SU(2)_L$ and $U(1)_Y$ gauge fields denoted by $W_\mu^a$ and $B_\mu$, respectively, and the Higgs boson $\Phi=(\phi_u~\,\phi_d)^T$ in the electroweak sector. The first generation of the left-handed $SU(2)_L$ lepton doublet field is denoted by $\ell_L\equiv(\nu_L, e^-_L)^T$ and the right-handed electron singlet field by $e^-_R$. Besides, we introduce the right-handed neutrino $\nu_R$ for which the Majorana mass term is given in the model in addition to the Dirac mass term composed together with the left-handed neutrino and the Higgs field.
The total Lagrangian is given by $\mathcal L_{\rm tot} =\mathcal L_{\rm SM} +\mathcal L_{\nu_R}$ where
\begin{align}
\label{LSM}
{\cal L}_{\rm SM} &=  -\frac{1}{4}W^a_{\mu\nu}W^{a\mu\nu}-\frac{1}{4}
F_{\mu\nu}F^{\mu\nu}+
     \left(D_\mu\Phi\right)^\dagger\left(D^\mu\Phi\right)
-\lambda\left(\Phi^\dagger\Phi-\frac{v_\Phi^2}{2}\right)^2  \nn
&\qquad + {\ell_L^\dagger}i\bar{\sigma}^\mu D_\mu\ell_L
+ e_R^\dagger i\sigma^\mu D_\mu e_R 
- h'\left(e_R^\dagger\Phi^\dagger\ell_L + {\ell_L^\dagger}\Phi e_R\right)\,,\\
\label{R neutrino sector}
{\cal L}_{\nu_R} &= \nu_R^\dagger i\sigma^\mu \partial_\mu \nu_R 
- \frac{1}{2} (\nu_R^c)^\dagger M^\ast_R \nu_R
- \frac{1}{2} \nu_R^\dagger M_R\nu_R^c 
-  h \left(\ell_L^\dagger i \tau_2 \Phi^\ast \nu_R 
- \nu_R^\dagger\Phi^T i \tau_2^{\dagger} \ell_L\right)\,.
\end{align}
Here, all spinors are given as two-component Weyl spinors and
the superscript $c$ on the right- and left-handed spinors denote the charge conjugation
such that $\nu_R^c=- i\sigma^2 \nu_R^\ast$ and $\nu_L^c= i\sigma^2 \nu_L^\ast$, respectively.
The covariant derivatives $D_\mu$ act on the fields $\Psi$ with hypercharge $Y$ such that
\begin{align}
D_\mu \Psi &= \left(\p_\mu  -i g\frac{\sigma^a}{2} W_\mu^a  -ig'\frac{Y}{2} B_\mu \right)\Psi\,,
\label{eq: covariant derivative on Phi}
\end{align}
where $g$ and $g'$ are the $SU(2)$ and $U(1)$ gauge couplings, respectively, and $\sigma^a$ are the Pauli matrices.
The field strengths of $W_\mu^a$ and $B_\mu$ are given respectively by
\begin{align}
W^a_{\mu\nu} & \equiv  \partial_\mu W^a_\nu-\partial_\nu W^a_\mu+ g\epsilon^{abc} W^b_\mu W^c_\nu \ , \\
F_{\mu\nu}   & \equiv  \partial_\mu B_\nu- \partial_\nu B_\mu\,.
\end{align}
The charge assignment for each particle is summarized in Table~\ref{standard model particle content}.
\begin{table}  
\begin{center}
\begin{tabular}{|c||c|c|} \hline
    field & $SU(2)_L$ & $Y$ \\ \hline \hline 
$\ell _L^T = (\nu _{e L}\,~  e_L)$ & \bf 2& $-1$ \\ \hline
$e_R$  & \bf 1 & $-2$  \\ \hline
$\nu_R$  & \bf 1 & $0$  \\ \hline
   $\Phi^T= (\phi_u\,~\phi_d) $  & \bf 2 & $1$   \\ \hline
   $W_\mu^a$  & \bf 3 & 0  \\ \hline
   $B_\mu$  & \bf 1 & 1  \\ \hline
  \end{tabular}
  \caption{Charge assignment for the particle contents}
  \label{standard model particle content}
  \end{center}
\end{table}

After the spontaneous symmetry breaking of $SU(2)_L\times U(1)_Y$ into $U(1)_{\rm em}$, i.e. the Higgs field obtains a non-vanishing expectation value $\langle\Phi\rangle= (0~~ v_\Phi/\sqrt{2})^T$, the gauge fields have mass terms characterized by $v_\Phi$.
As is conventional, we define the $Z$ boson and photon fields respectively by
\begin{align}
Z_\mu \equiv & \cos\theta_W W^3_\mu-\sin\theta_W B_\mu \,,  \label{eq: Z: WB} \\
A_\mu \equiv & \sin\theta_W W^3_\mu+\cos\theta_W B_\mu  \,,
\label{eq: A: WB}
\end{align}
where $\theta_W$ is the Weinberg angle which is related to the gauge couplings as
\begin{align}
&\cos\theta_W =\frac{g}{\sqrt{g^2+g'^2}}\,,&
&\sin\theta_W =\frac{g'}{\sqrt{g^2+g'^2}}\,.
\end{align}
Hereafter, we write $s_W\equiv \sin\theta_W$ and $c_W\equiv \cos\theta_W$ for simplicity.
The charged $W$ boson fields are given by
\begin{align}
W_\mu^\pm =\frac{1}{\sqrt{2}}(W_\mu^1 \mp i W_\mu^2)\,.
\end{align}
In the broken phase, the covariant derivative \eqref{eq: covariant derivative on Phi} is written in terms of the new basis as
\begin{align}
D_\mu\Psi  &= \left(\p_\mu  -\frac{i g}{\sqrt{2}} \sigma^+ W_\mu^+ - \frac{i g}{\sqrt{2}} \sigma^- W_\mu^-  - ig_Z\left(\frac{\sigma^3}{2}  - s_W^2 Q \right)Z_\mu - ie QA_\mu \right)\Psi\,,
\end{align}
where $\sigma^\pm=(\sigma^1\pm i\sigma^2)/2$ and $Q=\sigma^3/2 + Y/2$ is the electric charge defined through 
\begin{align}
g^2 \frac{\sigma^3}{2} -g'^2\frac{Y}{2} = g_Z^2\frac{\sigma^3}{2} - g'^2 Q\,.
\end{align}
We have also defined
\begin{align}
&g_Z= \sqrt{g^2+g'^2}=\frac{g}{c_W}\,,&
&e=\frac{gg'}{\sqrt{g^2+g'^2}}=g s_W\,.
\end{align}

We recast the Lagrangian for the gauge sector in the broken phase as
\begin{align}
\mathcal L^\text{Kin}_\text{gauge}
&=-\frac{1}{4} A_{\mu\nu} A^{\mu\nu} -\frac{1}{4}Z_{\mu\nu}Z^{\mu\nu} -\frac{M_Z^2}{2}Z_\mu Z^\mu  -\frac{1}{2}W^+_{\mu\nu} W^-{}^{\mu\nu} -M_W^2 W^+_\mu W^{-\mu}\,,\\[2ex]
\mathcal L^\text{Int}_\text{gauge}
&=iN_{\mu\nu}W^{+\mu} W^{-\nu}
- i W^+_{\mu\nu}N^\mu W^{-\nu}
+iW^-_{\mu\nu}N^\mu W^{+\nu}
- W^+_\mu W^{-\mu} N_\nu N^\nu
+  W^+_\mu W^-_\nu N^\mu N^\nu \nn
&\quad 
-\frac{g^2}{2}(W^-_\mu W^{-\mu}W^+_\nu W^{+\nu} - W^+_\mu W^{-\mu} W^-_\nu W^{+\nu})\,,
\end{align}
where the linear combination of $A_\mu$ and $Z_\mu$ has been defined as
\begin{align}
N_{\mu}=e A_{\mu} - g_Z c_W^2 Z_{\mu}\,.
\end{align}
The field strengths of $A_\mu$, $Z_\mu$ and $W_\mu^\pm$ are given respectively by
\begin{align}
&A_{\mu\nu} =\p_\mu A_\nu -\p_\nu A_\mu\,,&
&Z_{\mu\nu} =\p_\mu Z_\nu -\p_\nu Z_\mu\,,&
&W_{\mu\nu}^\pm =\p_\mu W^\pm_\nu -\p_\nu W^\pm_\mu\,,
\end{align}
while the field strength of $N_\mu$ is $N_{\mu\nu}=e A_{\mu\nu} - g_Z c_W^2 Z_{\mu\nu}$.
The gauge and the Higgs boson obtain the finite masses at the classical level as
\begin{align}
&M_W=\frac{1}{2}gv_\Phi\,,&
&M_Z= \frac{1}{2}g_Zv_\Phi\,,&
&M_H= \sqrt{2\lambda v_\Phi^2}\,.
\end{align}

Now, we focus on the $Z$-string configuration in the SM sector~\eqref{LSM}. 
To this end, we make the $Z$-string ansatz~\cite{Nambu:1977ag,Vachaspati:1992fi}: In cylindrical coordinates $(r,\theta,z)$, the gauge and Higgs fields
take the form
\begin{align}
&q Z_\theta= -n \zeta(r)\,,&
&\phi_d=  \frac{v_\Phi}{\sqrt{2}}f(r)e^{in\theta}\,,\nn
&Z_r=Z_z=Z_t=W_\mu^\pm =A_\mu=0\,,&
&\phi_u=0\,.
\label{eq: $Z$-string ansatz}
\end{align}
Here, we have defined $q \equiv g_Z /2$ and the winding number (or vorticity) $n$, and $\zeta(r)$ and $f(r)$ are the $Z$-string profile functions which are given as non-trivial solutions to the equations of motion for $Z_\mu$ and $\phi_d$:
\begin{align}
&f'' + \frac{1}{r}f'- \frac{n^2(1-\zeta)^2}{r^2}f - \beta f (f^2-1)=0\,,
\label{eq: equation of motion for f in classical level}
\\[2ex]
&\zeta'' - \frac{1}{r}\zeta' + 2(1-\zeta)f^2=0\,,
\label{eq: equation of motion for zeta in classical level}
\end{align}
with the boundary conditions $f(0)=\zeta(0)=0$ and $ f(\infty)=\zeta(\infty) = 1$. Here, we  have rescaled the radius $\tilde{r} =M_Z r/\sqrt{2}$, and omitted tilde for simple notation.
In Eq.~\eqref{eq: equation of motion for f in classical level}, we have defined
\begin{align}
\beta= \frac{8\lambda}{g_Z^2}=\frac{M_H^2}{M_Z^2}\,,
\end{align}
which is about $1.887$ in the SM. Here and hereafter, we use the prime as the derivative with respect to $r$. 
The exact solution to Eqs.~\eqref{eq: equation of motion for f in classical level} and \eqref{eq: equation of motion for zeta in classical level} is not known, so that we solve them numerically.

\subsection{Stability of \texorpdfstring{$Z$}{} string}
\label{sec: Stability of $Z$-string}
In general, the $Z$-string is not a stable object. This is in contrast to the ANO string whose stability is guaranteed by topology. The stability analysis for the $Z$-string has been performed in Refs.~\cite{James:1992zp,James:1992wb,Goodband:1995he}. In this section, we review its stability by following Ref.~\cite{Goodband:1995he}.

\subsubsection{Setup}
We first denote a set of fields by $\Phi=\{ \phi_u, \phi^*_u,  W_\mu^+,  W_\mu^-,  A_\mu , \phi_d,\phi_d^*,  Z_\mu\}$ and that of the background fields by $\bar\Phi=\{\bar \phi_u, \bar\phi^*_u, \bar W_\mu^+, \bar W_\mu^-, \bar A_\mu ,\bar \phi_d, \bar \phi_d^*, \bar Z_\mu\}$ which correspond to the configuration \eqref{eq: $Z$-string ansatz}.
We introduce an infinitesimal perturbation $\delta \Phi$ around the static background $\bar \Phi$ 
and substitute it into the full EOM for $\Phi=\bar \Phi + \delta \Phi$,
\begin{align}
\frac{\delta S}{\delta \Phi(x)}  =0 \, ,
\end{align}
from which we get
\begin{align}
\frac{\delta^2 S}{\delta\Phi \delta\Phi}\Big|_{\Phi=\bar\Phi} \, \delta \Phi=0 \label{eq:second-deriv}
\end{align}
at the leading order of the fluctuation $\delta \Phi$.
We have used that first derivative of the action vanishes since the background $\bar \Phi$ satisfies the static EOM.
In particular, those for $\bar\phi_d$ and $\bar Z^\mu$ read Eqs.~\eqref{eq: equation of motion for f in classical level} and \eqref{eq: equation of motion for zeta in classical level}, respectively.
This equation \eqref{eq:second-deriv} is decomposed into the four block diagonal parts
\begin{align}
&{\mathcal D}_1\pmat{\delta\phi_u \\\delta W_\mu^+}=0,&
&{\mathcal D}_2\pmat{\delta\phi_u^* \\\delta W_\mu^-}=0,&
&{\mathcal D}_3\pmat{\delta \phi_d\\\delta \phi_d^*\\\delta Z_\mu}=0,&
&{\mathcal D}_4\delta A_\mu=0\,.
\end{align}
The differential operators $\mathcal D_i$ correspond to the inverse propagators of fields around the background field $\bar\Phi$, so that it is convenient to use the gauge fixing action in order to remove zeromodes. To this end, we here employ
\begin{align}
S_{\rm gf}= \int \df^4x\,\mathcal L_{\rm gf}=\frac{1}{2}\int \df^4x\sum_{i=1}^4 |F_i|^2\,,
\end{align}
with the following $R_\xi$-gauge fixing functions 
\begin{subequations}
\begin{align}
F_1(W_\mu^+)&= \p^\mu\delta W_\mu^+ -i c_W^2 \bar Z^\mu \delta W_\mu^+ -\frac{ic_W}{\sqrt{2}}\bar\phi_d^*  \delta\phi_u\,,\\
F_2(W_\mu^-)&= \p^\mu \delta W_\mu^- + i c_W^2 \bar Z^\mu \delta W_\mu^- -\frac{ic_W}{\sqrt{2}}\bar\phi_d \delta\phi_u^*\,,\\
F_3(Z_\mu) &= \p^\mu \delta Z_\mu -\frac{i}{2}(\bar\phi_d\delta\phi_d^* -\bar\phi_d^* \delta\phi_d)\,,\\
F_4(A_\mu)&= \p^\mu \delta A_\mu\,.
\end{align}
\end{subequations}
These functions modify the differential operators $\mathcal D_i$ into $\tilde{\mathcal D}_i$. It turns out that the sector $(\delta \phi_d, \delta \phi_d^*, \delta Z_\mu, \delta A_\mu)$ always yields positive eigenvalues, whereas the sectors $(\delta\phi_u, \delta W_\mu^+)$ and its complex conjugate, namely
\begin{align}
&\tilde{\mathcal D}_1 \pmat{
\delta\phi_u \\
\delta W_\mu^+
}=0\,,&
\tilde{\mathcal D}_2 \pmat{
\delta\phi_u^* \\
\delta W_\mu^-
}=0\,,
\label{eq: eigenequation for phiu and W}
\end{align}
could contain unstable modes. Here, $\tilde{\mathcal D}_2$ can be obtained from $\tilde {\mathcal D}_1$, so that we concentrate on the $(\delta\phi_u, \delta W_\mu^+)$ sector.
We expand the gauge field $\delta W_\mu^\pm$ in eigenstates of spin operator $S_z$ so as to be $S_z W_\uparrow^\pm= + W_\uparrow^\pm$ and $S_z W_\downarrow^\pm= - W_\downarrow^\pm$. More specifically, one has
\begin{align}
&\delta W_\uparrow^\pm = e^{-i\theta} \left( W_r^\pm -\frac{i}{r}W_\theta^\pm \right)\,,&
&\delta W_\downarrow^\pm = (\delta W_\uparrow^{\pm})^\dagger\,.
\end{align}
To obtain the eigenequations, we perform the separation of spacetime variables such that
\begin{align}
&\delta \phi_u(x)= s_u(r) e^{i\ell \theta}e^{i\omega t}\,,
\label{eq: separation of spacetime variables for phiu}
\\[1ex]
&\delta W_\uparrow^+(x) = iw_{\uparrow,\ell}(r) e^{i(\ell-1-n)\theta} e^{i\omega t}\,,
\label{eq: separation of spacetime variables for Wup}
\\[1ex]
&\delta W_\downarrow^+(x) = iw_{\downarrow,\ell}(r) e^{i(\ell+1-n)\theta} e^{i\omega t}\,.
\label{eq: separation of spacetime variables for Wdown}
\end{align}
Inserting Eqs.~\eqref{eq: separation of spacetime variables for phiu}--\eqref{eq: separation of spacetime variables for Wdown} into the first equation in Eq.~\eqref{eq: eigenequation for phiu and W} leads to
\begin{align}
{\mathcal M}_1 \pmat{s_{u,\ell} \\[1ex] w_{\uparrow,\ell} \\[1ex] w_{\downarrow,\ell}} = \omega^2 \pmat{s_{u,\ell} \\[1ex] w_{\uparrow,\ell} \\[1ex] w_{\downarrow,\ell}}\,,
\label{eq: eigenequation for phiu WW}
\end{align}
where the stability matrix is given by
\begin{align}
\mathcal M_1 
=\pmat{
D_{1,\ell} && B_{1\uparrow,\ell} && B_{1\downarrow,\ell}\\[2ex]
B_{1\uparrow,\ell} && D_{\uparrow,\ell} && 0 \\[2ex]
B_{1\downarrow,\ell} && 0 && D_{\downarrow\,\ell}
}\,,
\label{eq: stability matrix of $Z$-string}
\end{align}
with the differential operators
\begin{subequations}
\begin{align}
&D_{1,\ell} = -\frac{\df^2 }{\df r^2} - \frac{1}{r}\frac{\df}{\df r} + \frac{(\ell- n \cos2\theta_W\zeta )^2}{r^2}  +2c_W^2f^2 + \beta (f^2-1)\,,
\label{eq: stability element phiudagger phiu}
\\[2ex]
&D_{\uparrow,\ell}= -\frac{\df^2 }{\df r^2} - \frac{1}{r}\frac{\df}{\df r}
 + \frac{(\ell-1 -n -2nc_W^2 \zeta )^2}{r^2} + 2c_W^2f^2 + 4nc_W^2 \frac{\zeta'}{r}\,,
 \label{eq: stability element Wp}
\\[2ex]
&D_{\downarrow,\ell}= -\frac{\df^2 }{\df r^2} - \frac{1}{r}\frac{\df}{\df r}+ \frac{(\ell + 1 -n -2nc_W^2 \zeta )^2}{r^2} + 2c_W^2f^2 - 4nc_W^2 \frac{\zeta'}{r}\,,
 \label{eq: stability element Wm}
\\[2ex]
&B_{1\uparrow,\ell}= 2c_W \left( f' -\frac{nf(1-\zeta)}{r} \right)\,,\\[2ex]
&B_{1\downarrow,\ell}=-  2c_W \left( f' + \frac{nf(1-\zeta)}{r} \right)\,.
\end{align}
\end{subequations}
Here, we have used the dimensionless quantities
\begin{align}
 &A_\mu= \frac{v_\Phi}{\sqrt{2}}\tilde A_\mu\,,&
 &W^\pm_\mu = \frac{v_\Phi}{\sqrt{2}} \tilde W^\pm_\mu\,,&
 &Z_\mu= \frac{v_\Phi}{\sqrt{2}} \tilde Z_\mu\,,&
 &\Phi =\frac{v_\Phi}{\sqrt{2}}\tilde \Phi\,,
 \label{eq: dimensionless quantities}
 \end{align}
and then have omitted tildes on them. Note that in this energy unit, the eigenvalues $\omega$ are given in the energy unit $M_Z/\sqrt{2}\equiv 1$ for which the Higgs mass is given as $M_H^2=2\beta$. 
Evaluating the eigenvalues of the stability matrix \eqref{eq: stability matrix of $Z$-string}, one can investigate the instability by exploring negative eigenvalues.
In the operators $D_{\uparrow,\ell}$ and $D_{\downarrow,\ell}$, the term $2c_W^2 f^2$ corresponds to the $W$ boson mass due to the non-vanishing Higgs background field $|\phi_d|=f(r)$.
The third terms in the operators $D_{i,\ell}$ are the angular momenta. We should especially note that the non-vanishing $Z$ boson field background $\bar Z_\theta=-\zeta(r)/q$ plays a role of an external ``$SU(2)$-magnetic field'', $\vec B_Z=\vec \p\times \vec {\bar Z}$.
Thus, due to the interaction between $\vec B_Z$ and the spin of the $W$ bosons ($S_z=\pm 1$), there exist the spin magnetic moments $\pm 4nc_W^2\zeta'/r$ in Eqs.~\eqref{eq: stability element Wp} and \eqref{eq: stability element Wm}. As we will see in \Cref{sec: Origin of instability}, these effects become one of origins of instability.

\subsubsection{Origin of instability}
\label{sec: Origin of instability}
We are interested especially in the case of $n=1$ and $\ell=0$ for which a negative eigenvalue $\omega^2<0$ is observed. Below, we analyze the eigenequation \eqref{eq: eigenequation for phiu WW} only in that case.

\begin{figure}
\centering
\includegraphics[width=10cm]{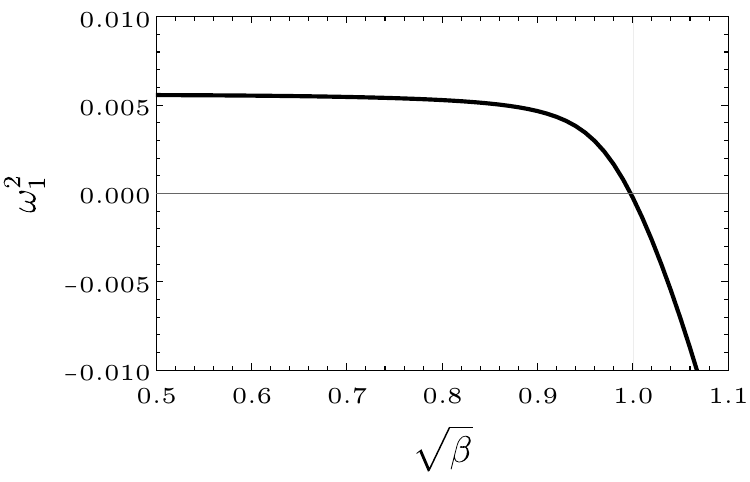}
\caption{
$\sqrt{\beta}$-dependence of the lowest eigenvalue of the stability matrix  \eqref{eq: stability matrix of $Z$-string} with $s_W^2=1$. Note $\beta\equiv 8\lambda/g_Z^2=M_H^2/M_Z^2$.
}
\label{fig:zero temperature eigenvalues for n=1 l=1 varying beta}
\end{figure}

We first consider the case of $c_W=0$ ($s_W=1$) for which the $Z$ boson field is identified with $U(1)_Y$ gauge field as can be seen from Eq.~\eqref{eq: Z: WB}: $Z_\mu = -B_\mu$. Therefore, the $Z$-string is regarded as a semilocal string~\cite{Vachaspati:1991dz}. The stability matrix \eqref{eq: stability matrix of $Z$-string} in this case becomes diagonal:
\begin{align}
\mathcal M_1 
&=
{\rm diag}\left(D_{1,0},~ D_{\uparrow,0},~ D_{\downarrow,0}\right)
\nn
&=\pmat{
-\frac{\df^2 }{\df r^2} - \frac{1}{r}\frac{\df}{\df r} + \frac{\zeta^2}{r^2} + \beta (f^2-1) && 0 && 0\\[2ex]
0 && -\frac{\df^2}{\df r^2} - \frac{1}{r}\frac{\df}{\df r}
 + \frac{4}{r^2} && 0 \\[2ex]
0 && 0 &&-\frac{\df^2 }{\df r^2} - \frac{1}{r}\frac{\df}{\df r}
}\,.
\label{eq: stability matrix in sin2=0}
\end{align}
Therefore, our task is to evaluate and to compare the eigenvalues of $D_{i,0}$ ($i=1,\uparrow,\downarrow$),
\begin{align}
&D_{1,0} s_{u,0}=\omega_1^2s_{u,0}\,,&
&D_{\uparrow,0} w_{\uparrow,0} = \omega_\uparrow^2w_{\uparrow,0}\,,&
&D_{\downarrow,0} w_{\downarrow,0} = \omega_\downarrow^2w_{\downarrow,0}\,.
\label{eq: separeted eigenequations}
\end{align}
Because the fluctuation of the $W$ bosons are decoupled from the background,
the operators $D_{\uparrow,0}$ and $D_{\downarrow,0}$ are independent of both $\beta$ and the profile of the string solutions, $f(r)$ and $\zeta(r)$. Indeed, the second and third equations in Eq.~\eqref{eq: separeted eigenequations},
\begin{align}
&\left(-\frac{\df^2}{\df r^2} - \frac{1}{r}\frac{\df}{\df r}+ \frac{4}{r^2} \right) w_{\uparrow,0}  = \omega_\uparrow^2 w_{\uparrow,0} \,,&
& \left( -\frac{\df^2 }{\df r^2} - \frac{1}{r}\frac{\df}{\df r} \right) w_{\downarrow,0}=\omega_{\downarrow}^2 w_{\downarrow,0}\,.
\end{align}
are classes of the Helmholtz equation which yields positive eigenvalues, $\omega_\uparrow^2>0$ and $\omega_{\downarrow}^2>0$. Note that the solutions to these equations are given by
\begin{align}
w_{\uparrow,0} (r) &= c_1 J_2(\omega_\uparrow r)+c_2 Y_2(\omega_\uparrow r)\,,&
w_{\downarrow,0}(r) &= c_3 J_0(\omega_\downarrow r)+c_4 Y_0(\omega_\downarrow r)\,,
\end{align}
where $J_n(x)$ and $Y_n(x)$ are the Bessel functions of the first and second kinds, respectively, and $c_i$ ($i=1,\ldots,4$) are constant.
On the other hand, the eigenvalue of $D_{1,0}$ explicitly depends on $\beta$ and the profile functions. In Fig.~\ref{fig:zero temperature eigenvalues for n=1 l=1 varying beta}, the $\sqrt{\beta}$-dependence of the lowest eigenvalues of the first equation in Eq.~\eqref{eq: separeted eigenequations} is displayed. Here and hereafter, to perform numerical calculations, we set the boundary condition $s_{u,0}(r_\text{max})=0$ with $r_\text{max}=50$ instead of $r_\text{max}=\infty$.
We see that $\omega_1^2$ is positive for $\sqrt{\beta} < 1$, while for $\sqrt{\beta}> 1$, it turns to negative. 
This is understood as that the Higgs potential gives a tachyonic contribution proportional to $\beta$,
which is balanced at $\beta=1$ with the positive contribution from the kinetic term.
This agrees with the stability analysis in semilocal string~\cite{Hindmarsh:1991jq,Achucarro:1992hs}.

We next turn to the case $\beta$ to be small
so that 
the tachyonic contribution from Higgs potential becomes insignificant and hence
the instability comes only from the gauge sector.
Especially when $s_W=0$, the $Z$ boson is identified with $W_\mu^3$, while the electromagnetic field $A_\mu$ is just the $U(1)_Y$ field, $B_\mu$. In this case, the so-called ``$W$-condensation'' is clearly observed as an origin of instability. 
To explain $W$-condensation, let us here consider the ``$SU(2)_L$-magnetic background field'' defined by $B_{Z,i}=\epsilon_{ijk}\p_j \bar Z_k$ which couples to the $W$ bosons with the coupling $g_Zc_W^2$. Assuming that $\vec B_{Z}=(0,0,B_Z)$, the $x$- and $y$-direction momenta $p_x^2+p_y^2$ of the charged $W$ bosons, $W_\mu^{\pm}$, are discretized as the Landau levels such that their energy dispersion is given by 
\begin{align}
E^2=(2l+1)g_Zc_W^2 B_Z+p_z^2+M_W^2+2g_Zc_W^2B_Z S_z\,,
\label{eq: Landau levels}
\end{align}
where $p_z$ is the momentum along with the $z$-direction and $l$ are the discrete positive integer numbers, $l=0,1,2,\ldots$, denoting the Landau levels. Here, the last term in Eq.~\eqref{eq: Landau levels} corresponds to the magnetic moment between the spin $S_z=\pm 1$ of the $W$ bosons and $B_Z$. In the lowest energy of the spin $S_z=-1$ case, namely $p_z=0$ and $l=0$ for which one has $E^2=M_W^2-g_Zc_W^2B_Z$, the energy could be negative when
\begin{align}
B_Z> \frac{M_W^2}{g_Zc^2_W}\,.
\label{eq: BZ bound}
\end{align}
This instability is called ``$W$ condensation''~\cite{Ambjorn:1989sz,Vachaspati:1992jk,Achucarro:1993bu,Perkins:1993qz}. Thus, as the background field $B_Z$ is stronger, the $W$ bosons tend to form condensation at the Landau level $l=0$. 
This is confirmed by analyzing the stability matrix \eqref{eq: stability matrix of $Z$-string} numerically.

\begin{figure}
\centering
\includegraphics[width=0.48\columnwidth]{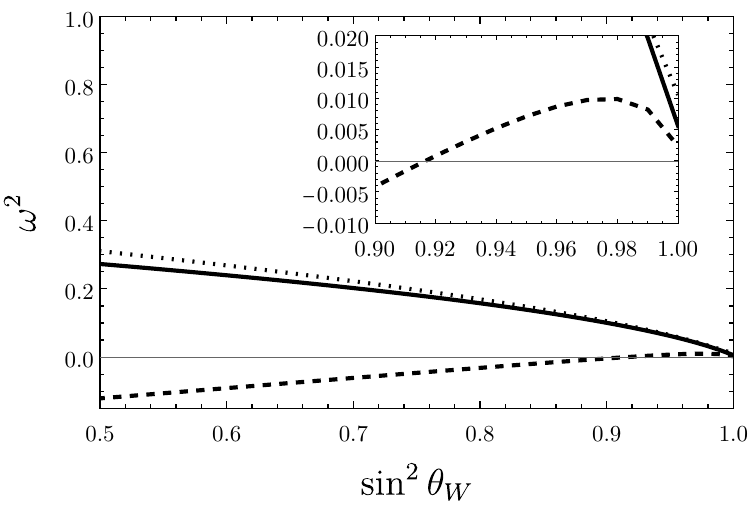}
\hspace{2ex}
\includegraphics[width=0.48\columnwidth]{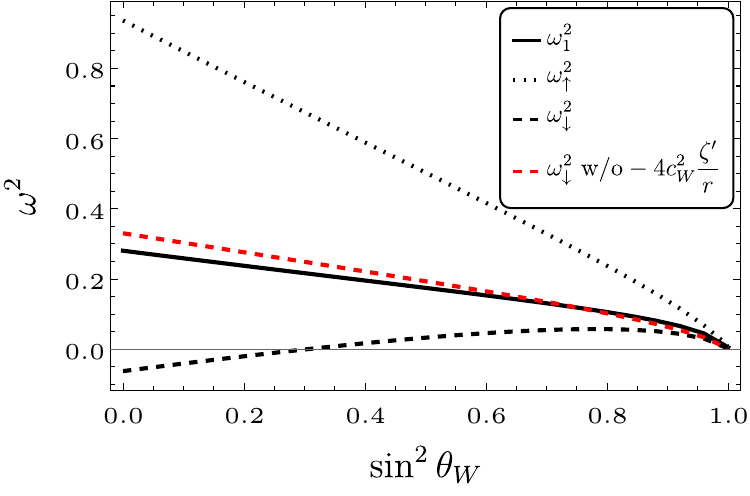}
\caption{
$\sin^2\theta_W$-dependence of the lowest three eigenvalues (black dashed, solid, and dotted lines) of stability matrices of Eq.~\eqref{eq: stability matrix of $Z$-string} with $\beta=0.01$ (left) and of the approximated version in Eq.~\eqref{eq: approxmatecd stability matrix} with $\beta=0.01$ (right).
The red dashed line in the right panel indicates the lowest eigenvalue for the stability matrix in which the term $4 c_W^2 \frac{\zeta'}{r}$ is eliminated by hand.
}
\label{fig:zero temperature eigenvalues for n=1 l=1 varying sin}
\end{figure}

In terms of the stability matrix \eqref{eq: stability matrix of $Z$-string}, the last term in the operator \eqref{eq: stability element Wm}, namely $-4c_W^2\zeta'/r$, corresponds to the spin magnetic moment which causes an instability due to $W$-condensation.
To see this, let us first consider the approximated stability matrix
\begin{align}
\mathcal M_1 
\approx
{\rm diag} \Bigg(&
-\frac{\df^2 }{\df r^2} - \frac{1}{r}\frac{\df}{\df r} + \frac{( \cos2\theta_W\zeta )^2}{r^2}  +2c_W^2f^2 + \beta (f^2-1),\nn[1ex]
&\qquad
-\frac{\df^2 }{\df r^2} - \frac{1}{r}\frac{\df}{\df r}
 + \frac{4(1 + c_W^2 \zeta )^2}{r^2} + 2c_W^2f^2 + 4c_W^2 \frac{\zeta'}{r},\nn[1ex]
&\qquad\qquad -\frac{\df^2 }{\df r^2} - \frac{1}{r}\frac{\df}{\df r}+ \frac{4c_W^4 \zeta^2}{r^2} + 2c_W^2f^2 - 4c_W^2\frac{\zeta'}{r}
\Bigg)\,,
\label{eq: approxmatecd stability matrix}
\end{align}
and then investigate eigenvalues of the each diagonal part.
The right side panel of Fig.~\ref{fig:zero temperature eigenvalues for n=1 l=1 varying sin} depicts each eigenvalue of Eq.~\eqref{eq: approxmatecd stability matrix}. Here, the eigenvalues are denoted as the same as of Eq.~\eqref{eq: separeted eigenequations}. We can see that $D_{1,0}$ and $D_{\uparrow,0}$ ($\omega_1^2$ and $\omega_{\uparrow}^2$) yield positive eigenvalues, while $D_{\downarrow,0}$ produces negatives $\omega_{\downarrow}^2$ for $\sin^2\theta_W<0.3$. When we eliminate $-4c_W^2\zeta'/r$ from $D_{\downarrow,0}$ by hand, its eigenvalues become positive for all range of $\sin^2\theta_W$ as shown by the red dashed line in the right panel of Fig.~\ref{fig:zero temperature eigenvalues for n=1 l=1 varying sin}. 
On the other hand, the left panel of Fig.~\ref{fig:zero temperature eigenvalues for n=1 l=1 varying sin} shows its three lowest eigenvalues of the stability matrix \eqref{eq: stability matrix of $Z$-string} without any approximation for $\beta=0.01$. 
One of eigenvalues (the dashed line in Fig.~\ref{fig:zero temperature eigenvalues for n=1 l=1 varying sin}) takes a negative value for $\sin^2\theta_W\lsim 0.92$.
One can see that qualitative tendency agrees with the approximated one (right panel).

\begin{figure}
\centering
\includegraphics[width=13cm]{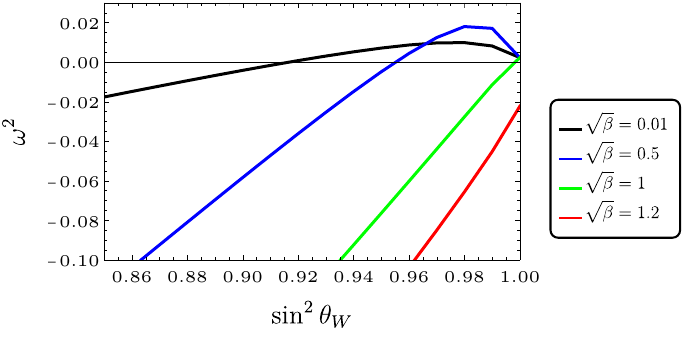}
\caption{
$\sin^2\theta_W$-dependence of the lowest eigenvalue of stability matrix~\eqref{eq: stability matrix of $Z$-string} with different values of $\sqrt{\beta}$. Note $\beta\equiv 8\lambda/g_Z^2=M_H^2/M_Z^2$.
}
\label{fig:zero temperature eigenvalues for n=1 l=1 varying sin and beta}
\end{figure}

We show the dependence of the lowest eigenvalue of stability matrix \eqref{eq: stability matrix of $Z$-string} on $s_W^2$ and $\sqrt{\beta}$ in Fig.~\ref{fig:zero temperature eigenvalues for n=1 l=1 varying sin and beta}. Note that the $\sqrt{\beta}$-dependence of $\omega^2$ at $s_W^2=1$ agrees with that in Fig.~\ref{fig:zero temperature eigenvalues for n=1 l=1 varying beta}.
The larger values of $\sqrt{\beta}$ are, the more negative the lowest eigenvalue tends to be. In particular, for $\sqrt{\beta}> 1$, no value of $s_W^2$ yielding positive eigenvalues exists, which indicates that there is no stabilized region.

\begin{figure}
\centering
\includegraphics[width=10cm]{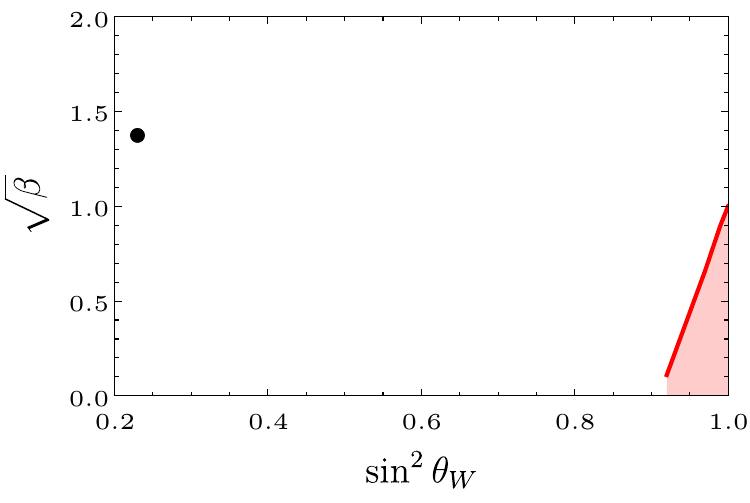}
\caption{
The $Z$-string is stabilized within the red region on $s_W^2$-$\sqrt{\beta}$ plane. The black point denotes the experimental values $s_W^2=0.23$ and $\sqrt{\beta}\simeq 1.37$.
}
\label{fig:stable region}
\end{figure}

To summarize, the stable region on $s^2_W$-$\sqrt{\beta}$ plane is displayed in Fig.~\ref{fig:stable region}. Only in the triangle region, the $Z$-string can stably exist in sense that it is robust under small perturbations. Obviously, the experimental values, $s_W^2=0.23$ and $\sqrt{\beta}\simeq 1.37$ (which is shown by the black point in Fig.~\ref{fig:stable region}), are out of the stabilized region, so that the $Z$-string in the SM is an unstable object.

In the following, we discuss the impact of fermion zeromodes on the $Z$-string~\cite{Starkman:2000bq,Starkman:2001tc} as a mechanics of its stabilization.

\subsubsection{Stabilization mechanism due to fermion zeromodes}

Let us here briefly explain a fate of fermion zeromodes on the electroweak string. We consider the doublet fields $\Psi=(\nu, e)^T,~(u,d)^T$, etc. on the electroweak string background. Note that zeromodes are degenerate energy eigenvalues. Solving the equations of motion (i.e., the Dirac equations) on the electroweak string background, one can see that the up and down components in $\Psi$ move in parallel and anti-parallel to the electroweak string flux, respectively~\cite{Earnshaw:1994jj,Garriga:1994wb}.

\begin{figure}[tbp]
\centering
\includegraphics[width=0.37\textwidth]{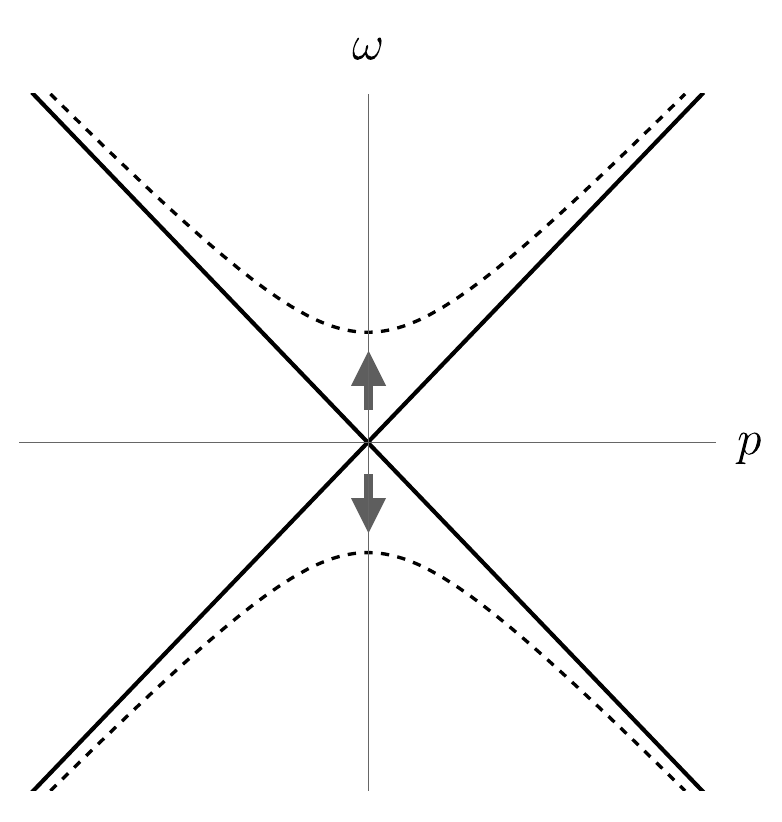}
\hspace{4mm}
\includegraphics[width=0.37\textwidth]{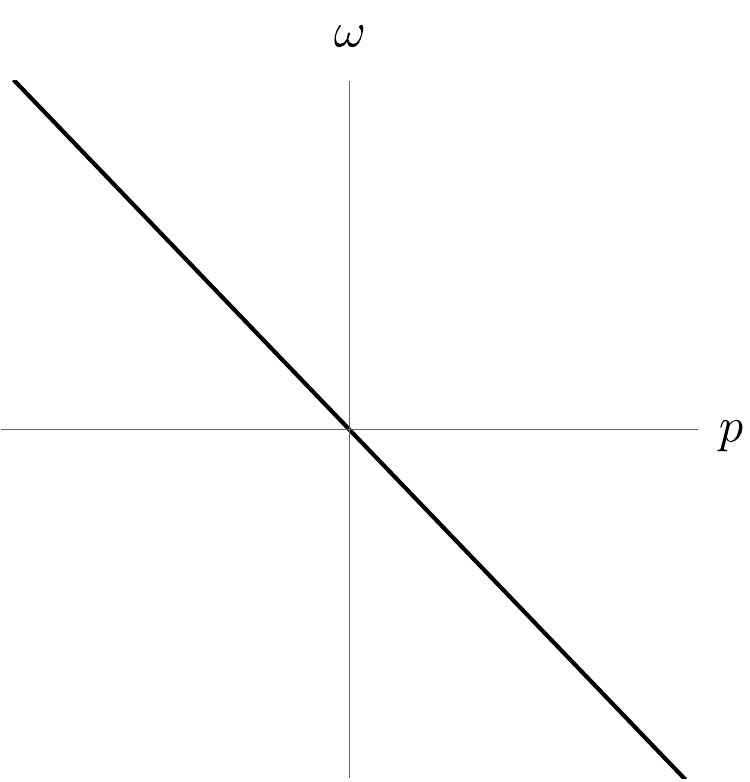}
\caption{
Illustration for two dimensional spectrum of fermion zeromodes in the string background.
If the zeromodes appear as pairs of right and left-mover Weyl fermions (solid line in left panel),
their degeneracy can be resolved by some perturbation,
resulting in massive Dirac fermions in two dimensions (dashed line in left panel).
In the sense of the Dirac index, this configuration is in the same topological sector as that of the vacuum (no zeromodes).
On the other hand, if there exists a zeromode with single chirality, e.g., only left-mover,
the zeromode cannot be made massive nor removed by any continuous perturbation (right panel).
This string configuration might be regarded as in a different topological sector from that of the vacuum
and is conjectured to be stable in Ref.~\cite{Starkman:2001tc}.
}
\label{fig:zeromode_spectrum}
\end{figure}

In this sense, when the electroweak string flux is in the $z$ direction, the up component is called a ``right-mover'' as proportional to $e^{-i\omega t+ipz}$ with a frequency $\omega$ and a momentum $p$, 
while the down one is  a ``left-mover'' as proportional to $e^{-i\omega t-ipz}$. 
For the ANO string with a single fermion, 
only one zeromode solution appears as a Weyl fermion in two dimensional sense~\cite{Jackiw:1981ee}
and cannot be removed by any continuous deformation of the configuration.
See the right panel in Fig.\ref{fig:zeromode_spectrum}.
However, this does not happen in the case of the electroweak string which is not topological: 
The string necessarily couples to a fermion doublet $\Psi$, 
which provides both of the left-mover and right-mover Weyl fermions as a pair to form a two dimensional Dirac fermion.
When the electroweak string is slightly perturbed with a parameter $\epsilon$, 
the degeneracy of the zeromode eigenvalues is resolved 
so that the Dirac fermion obtains finite masses of order of $\epsilon^2$.
(See the solid and dashed lines in the left panel in Fig.\ref{fig:zeromode_spectrum} for the unperturbed and perturbed spectrum.)
This is shown by the standard perturbation theory as
the correction to the zero energy eigenvalues at the first-order perturbation of order $\epsilon$ vanishes, i.e., $E^{(1)}=0$ 
and the energy eigenvalues are corrected by $E^{(2)}$ of the second order of $\epsilon$.
It has been argued in Refs.~\cite{Naculich:1995cb,Kono:1995xp,Liu:1995at} that fermions make the infinitely long electroweak string unstable due to the Dirac sea 
in which all negative energy eigenvalues should be summed up to give a negatively divergent contribution to $E^{(2)}$.
Thus, it seems that the string cannot be stabilized under any perturbation.

Now, we turn to see the stabilization mechanism of the electroweak string in Ref.~\cite{Starkman:2001tc} due to the fermion.
To this end, we assume that, in some extension of the neutrino sector, the up-component fields in $\Psi$ do not have zeromodes
while the down-component fields still keep zeromodes.
In such a situation, the zeromode-energy eigenmodes cannot be made massive in the same way as described above and thus survive under any continuous deformation
(see the right panel in Fig.\ref{fig:zeromode_spectrum}).
Then, the discrepancy of the number of zeromodes between the up and down components leads to the non-vanishing topological number in the sense of the index theorems.
Hence, the electroweak string is a topologically distinct object from the vacuum.
This implies that the string cannot decay by any continuous deformation.
The main scope of this paper is to revisit this mechanism in several models of extensions in the neutrino sector.

\section{Number of zeromodes on vortices and its relation with momentum space topology}
\label{sec:momentum_space_topology_1d}

Now, we are ready to discuss the fermion zeromodes that appear on the $Z$-string. 
To this end, we first present a convenient formalism based on the topological invariant defined in the momentum space (and real space),
originally introduced in Refs.~\cite{Zubkov:2016llc,Volovik:2016mre}.
In this section, we come back to the dimensionful unit for the coordinate $r$.

\subsection{Topological invariant for the number of zeromodes}

We consider a Green's function $\mathcal{G}(x^\mu,p_\mu)$ of a fermion field in a background of vortex strings, located on the $z$ axes. Due to the translational invariance with respect to the $z$ and $t$ directions, the derivatives with respect to $z$ and $t$ in the Green's function can be replaced by $p_3$ and $p_4(\equiv i p_0)$, 
while in the $x$ and $y$ directions it contains $x$, $y$, $\partial_x$, and $\partial_y$ explicitly.

Here, we introduce the following topological quantity:
\begin{equation}
\label{145951_8Mar21}
 \mathcal{N}_1 \equiv \frac{1}{2\pi i}\int_C \mathrm{Tr} \left[\mathcal{G}(p_3,p_4) \df \mathcal{G}(p_3,p_4)^{-1}\right]\,,
\end{equation}
where $\mathcal{G}(p_3,p_4)$ is the Green's function with $p_3$ and $p_4$ being $c$-numbers as stated above.
For instance, one may give $\mathcal{G}^{-1}=\gamma^\mu\hat{p}_\mu + M(\hat{x},\hat{y})|_{\hat{p}_3 = p_3, \, \hat{p}_4=p_4}$.
The integration contour $C$ is a circle with infinitesimal radius embracing $p_3=p_4=0$ on $p_3$-$p_4$ plane. The trace is taken over the function space of $x$ and $y$ (on which $\partial_x$ and $\partial_y$ act) and the spinor indices.
The fact that $\mathcal{N}_1$ is a topological invariant is understood immediately below.

To make the physical meaning of $\mathcal{N}_1$ clearer,
we start with the fact that $\mathcal{G}^{-1}$ can be expressed as
\begin{equation}
 \mathcal{G}^{-1} = (i p_4 + \hat{\mathcal{H}})\gamma^0 \,,
\end{equation}
where $\hat{\mathcal{H}}$ corresponds to the one-particle Hamiltonian in the background of the strings. Since it is Hermitian and commutes with $\hat{p}_3$ and $\hat{p}_4$, 
we can diagonalize it independently of $p_3$ and $p_4$.
Thus if one denotes the $n$-th eigenstates of $\hat{\mathcal{H}}$ by $\ket{n}$,
the quantity \eqref{145951_8Mar21} is given as follows:
\begin{align}
 \mathcal{N}_1 &= \frac{1}{2\pi i}\sum_n \bra{n}\int_C ~\mathcal{G}(p_3, p_4) \df \mathcal{G}(p_3,p_4)^{-1} \ket{n}\nn
&= \frac{1}{2\pi i}\sum_n \int_C ~G_n(p_3,p_4) \df G_n(p_3,p_4)^{-1}  , 
\label{eq: N1 in c number}
\end{align}
with 
\begin{equation}
 G_n^{-1} \equiv i p_4 + \mathcal{E}_n(p_3) \, ,
\end{equation}
where $\mathcal{E}_n$ is the eigenvalue of $\hat{\mathcal{H}}$ for the state $\ket{n}$.
From this expression, 
one can find that $\mathcal{N}_1$ is a topological invariant.
For each eigenstate $\ket{n}$, \footnote{For continuous spectrum, the summation over $n$ should be replaced with integration over continuous eigenvalues.}
the integration gives the winding number from the circle $C$ on $p_3$-$p_4$ plane to the complex phase of $G_n$,
which is characterized by the elements of the first homotopy group $\pi_1(U(1))=\mathbb{Z}$.
Note here that we have assumed that there is no massless particle in the bulk,
which corresponds to a continuous spectrum starting from zero in two dimensions.
Such a massless particle makes the integrand in Eq.~\eqref{145951_8Mar21} singular with a branch cut starting from the origin on the original Lorentzian $p_0$-$p_3$ plane 
which leads to an obstacle to perform the calculation.
Thus, the argument presented here is not applicable to the SM since the neutrinos are massless.

Defining a complex variable $z \equiv p_3 + i p_4$,
we can rewrite Eq.~\eqref{eq: N1 in c number} on the complex plane as
\begin{equation}
 \mathcal{N}_1 = \frac{1}{2\pi i}\sum_n \int_C \df z \frac{\partial}{\partial z}\log G_n^{-1}(p_3,p_4)  \, .\label{eq:N1-complex-integral}
\end{equation}
Note that the integration picks up residues of poles at $p_3=p_4=0$, and thus vanishes for such $n$ that $\mathcal{E}_n \neq 0$ at $p_3=p_4=0$. These are massive modes in two dimensions.\footnote{Massive modes provide poles or cuts on $p_0$-$p_3$ plane, while they do not on $p_3$-$p_4$ plane thanks to the Wick rotation.}
Therefore, only modes with vanishing $\mathcal{E}_n$ at $p_3=p_4=0$ (just corresponding to massless modes) can contribute to the integration. For such zeromodes, denoting $\mathcal{E}_n \simeq \lambda_n p_3 + \mathcal{O}(p_3^2)$,
we have
\begin{equation}
 \mathcal{N}_1 = \frac{1}{2\pi i}\sum_n \int_C \df z \frac{\partial}{\partial z}\log (i p_4 + \lambda_n p_3 ) \, ,
\end{equation}
which means that $ \mathcal{N}_1$ takes the sign of $\lambda_n$ for each massless mode.
Because $\lambda_n$ determines the two-dimensional chirality of the massless modes, this is nothing but the Dirac index in two dimensions, namely,
\begin{equation}
 \mathcal{N}_1 = n_+ - n_- \, ,
 \label{eq: total vorticity}
\end{equation}
where $n_+$ and $n_-$ are the numbers of the right- and left-mover modes on the vortex, respectively.

For later use, it is convenient to write the quantity \eqref{145951_8Mar21} in terms of the Wigner transform of the Green's function. Here, the Wigner transformed Green's function is defined by
\begin{equation}
 \tilde{G}(\bm{R},\bm{p},p_3,p_4) \equiv 
\int \df^2 r_R ~ e^{-i \bm{p}\cdot \bm{r}_R} 
\left\langle\bm{r}_1\right|\mathcal{G}(p_3, p_4) \left|\bm{r}_2\right\rangle \, ,
\end{equation}
where $\bm{r}_1$ and $\bm{r}_2$ are vectors in $x$-$y$ position space 
with $\bm{R}\equiv(\bm{r}_1+\bm{r}_2)/2$ and $\bm{r}_R\equiv\bm{r}_1-\bm{r}_2$ and $\bm{p}$ denotes a vector in the momentum $p_1$-$p_2$ space.
The bra and ket are eigenfunctions in the position space.
By definition, the following identity holds:
\begin{equation}
 \int \df^2 R \int \frac{\df^2 p}{(2 \pi)^2} ~\tilde{G}(\bm{R},\bm{p},p_3,p_4) = \mathrm{Tr} ~\mathcal{G}(p_3, p_4)\,.
\end{equation}
Thus, the Wigner transformation can be regarded as a map from an operator including the canonical operators $\hat{x}$, $\hat{p}$ 
into a \textit{classical} distribution function defined on the phase space $(x,p)$.

By using the Wigner transformations for $\mathcal{G}$ and $\mathcal{G}^{-1}$, which are denoted by $\tilde{G}$ and $\mathcal{Q}$ hereafter, respectively,
Eq.~\eqref{145951_8Mar21} reads
\begin{equation}
 \mathcal{N}_1 = \frac{1}{2\pi i}\int \df^2 R \int \frac{\df^2 p}{(2 \pi)^2} \int_C 
\mathrm{Tr}_D \left[\tilde{G}(\bm{R},\bm{p},p_3,p_4) \df \mathcal{Q}(\bm{R},\bm{p},p_3,p_4)\right] \,,
\end{equation}
where $\mathrm{Tr}_D$ denotes the trace acting on the spinor and other internal indices and $\mathcal{Q}$ is determined by the relation
\begin{align}
1& = \mathcal{Q}(\bm{R},\bm{p},p_3,p_4) \ast \tilde{G}(\bm{R},\bm{p},p_3,p_4) \nn
& \equiv\mathcal{Q}(\bm{R},\bm{p},p_3,p_4) 
\exp\left[{\frac{i}{2}(\overleftarrow{\partial_R} \cdot \overrightarrow{\partial_p} - \overleftarrow{\partial_p} \cdot\overrightarrow{\partial_R})} \right]
\tilde{G}(\bm{R},\bm{p},p_3,p_4) \,,
\label{153135_8Mar21}
\end{align}
which follows from the Wigner transformation of the identity $\mathcal{G} \mathcal{G}^{-1}=1$.
Based on the derivative expansion of $\bm{R}$ and $\bm{p}$,
Eq.~\eqref{153135_8Mar21} is solved iteratively as
\begin{align}
\tilde{G} (\bm{R},\bm{p},p_3,p_4) \equiv \tilde{G}^{(0)} + \tilde{G}^{(1)} + \tilde{G}^{(2)} + \cdots
\end{align}
with
\begin{align}
\tilde{G}^{(0)}(\bm{R},\bm{p},p_3,p_4) &= \mathcal{Q}(\bm{R},\bm{p},p_3,p_4) ^{-1} \, , \\
\tilde{G}^{(1)}(\bm{R},\bm{p},p_3,p_4) &= - \frac{i}{2} \tilde{G}^{(0)}  \left[\partial_{\bm{R}} \mathcal{Q} \cdot \partial_{\bm{p}} \tilde{G}^{(0)} - \partial_{\bm{p}}\mathcal{Q}\cdot \partial_{\bm{R}}\tilde{G}^{(0)} \right] \, , \\
\tilde{G}^{(2)}(\bm{R},\bm{p},p_3,p_4) &= - \frac{i}{2} \tilde{G}^{(0)}  \left[\partial_{\bm{R}} \mathcal{Q} \cdot \partial_{\bm{p}} \tilde{G}^{(1)} - \partial_{\bm{p}}\mathcal{Q}\cdot \partial_{\bm{R}}\tilde{G}^{(1)} \right]  \nonumber \\
& \hspace{2em} - \frac{1}{2}\left(\frac{i}{2}\right)^2 \left[
\partial_{R_i}\partial_{R_j} \mathcal{Q} \, \partial_{p_i} \partial_{p_j} \tilde{G}^{(0)} 
+ \partial_{p_i}\partial_{p_j} \mathcal{Q} \, \partial_{R_i} \partial_{R_j} \tilde{G}^{(0)}  \right. \nonumber \\
& \hspace{7em} \left. - 2 \partial_{p_i}\partial_{R_j} \mathcal{Q} \, \partial_{R_i} \partial_{p_j} \tilde{G}^{(0)} 
\right]\,.
\end{align}
This formula allows us to rewrite eventually the topological invariant $\mathcal{N}_1$ in the form of
\begin{align}
 \mathcal{N}_1 & = \frac{2}{5 ! (2\pi)^3 i} \int \mr{Tr}_D
\left[
\tilde{G}^{(0)} \df (\tilde{G}^{(0)} )^{-1} \wedge
\df \tilde{G}^{(0)} \wedge
\df (\tilde{G}^{(0)})^{-1} \wedge
\df \tilde{G}^{(0)} \wedge
\df (\tilde{G}^{(0)})^{-1}
\right] \nn
& = \frac{2}{5 ! (2\pi)^3i} \int \mr{Tr}_D
\left[
\tilde{G}^{(0)} \df (\tilde{G}^{(0)})^{-1} 
\right] ^5\,,
\label{164812_9Mar21}
\end{align}
where the integration is taken over the surface $C \otimes R^2 \otimes R^2$;
$C$ is the circle embracing $p_3=p_4=0$ in $p_3$-$p_4$ space,
the first $R^2$ corresponds to $x$-$y$, and
the second $R^2$ corresponds to $p_1$-$p_2$ space.
See Refs.~\cite{Silaev:2010za,Zubkov:2016llc}.
Now, one can see that $\mathcal{N}_1$ is expressed by the winding number characterizing maps from the five-dimensional space to $GL(N,\mathbb{C})$ with $N$ being the size of the matrix $\tilde{G}^{(0)}$.
Note that there are no singularities of $\tilde{G}$ in the entire six dimensional space other than the massless poles located on the origin.
Due to the topological nature, 
we can rotate the integration surface into $C' \otimes R^4$
keeping the massless poles included inside the integral region,
where $C'$ is a circle embracing $x=y=0$ in $x$-$y$ space
and $R^4$ corresponds to the momentum space $p^1,\cdots, p^4$.
Furthermore, 
as long as the mass gap is not closed in the bulk, or equivalently, 
a new singularity appears inside the integration circle $C$ in Eq.~\eqref{eq:N1-complex-integral},
this quantity is topologically protected under any perturbation of the model.
In the following, we calculate $\mathcal{N}_1$ \eqref{164812_9Mar21} with the integration surface $C' \otimes R^4$ for some benchmark cases.

\subsection{Dirac fermion on Abrikosov-Nielsen-Olesen vortex}
\label{sec: Dirac fermion on Abrikosov-Nielsen-Olesen vortex}
As an exercise, we start with a demonstration for counting the number of zeromodes with Eq.~\eqref{164812_9Mar21} in case of a Dirac fermion in the presence of the (global) ANO vortex. In this system, there exists a zeromode~\cite{Jackiw:1981ee}, so that the topological invariant \eqref{164812_9Mar21} has to entail nonzero value. 

The Green's function for the fermion field in the ANO vortex background is given as
\begin{equation}
 \mathcal{G}^{-1} = \gamma^i \hat{p}_i +\gamma^a p_a - m_D f(r) e^{i n\gamma_5 \theta}\,,
 \label{eq: Green's function of NO string}
\end{equation}
where $m_D$ is the Dirac fermion mass, $n$ is the vorticity, $x+i y = r e^{i\theta}$, $f(r)$ is the profile function satisfying $f(0)=0$ and $f(\infty)=1$, and the indices run as $i=1,2$, and $a=3,0$
(with $p_0=-ip_4$).
Note that $\gamma^i$ and $\gamma^0$ are anti-hermitian and hermitian, respectively.
The Wigner transformation for the Green's function \eqref{eq: Green's function of NO string} is computed explicitly as 
\begin{align}
& \left(\tilde{G}^{(0)} \right)^{-1} 
=\mathcal{Q} (\bm{R},\bm{p},p_3,p_4)  \nn
&\quad = \int \df^2r_R ~e^{ -i \bm{p \cdot \bm{r}_R}}
\left\langle\bm{R}+\frac{\bm{r}_R}{2}\right|
\left(\gamma^i \hat{p}_i +\gamma^a p_a  - m_D f(r) e^{i n\gamma_5 \theta} \right)
\left| \bm{R}-\frac{\bm{r}_R}{2}\right\rangle \nn
&\quad
= \int \df^2r_R ~e^{ -i \bm{p \cdot \bm{r}_R}} 
\left(\int \frac{\df^2 q}{(2\pi)^2}
\left\langle\bm{R}+\frac{\bm{r}_R}{2}\right| \gamma^i \hat{p}_i +\gamma^a p_a  \Big|q\Big\rangle \Big\langle q\left|\bm{R}-\frac{\bm{r}_R}{2}\right\rangle 
- \delta^2(\bm{r}_R) m_D f(r) e^{i n\gamma_5 \theta} \right)\nn
&\quad
= \int \df^2r_R \int \frac{\df^2 q}{(2\pi)^2} ~e^{ -i (\bm{p}-\bm{q}) \cdot \bm{r}_R}~
(\gamma^i q_i +\gamma^a p_a )
- \int \df^2r_R ~e^{ -i \bm{p \cdot \bm{r}_R}} ~\delta^2(\bm{r}_R) m_D f(r) e^{i n\gamma_5 \theta} \nn
&\quad
= \gamma^\mu p_\mu - m_D f(r) e^{i n\gamma_5 \theta} \,.
\end{align}
For later use, we rewrite this as
\begin{equation}
 \left(\tilde{G}^{(0)} \right)^{-1} =  e^{i n\gamma_5 \frac{\theta}{2}} \left(\gamma^\mu p_\mu - m_D f(r) \right) e^{i n\gamma_5 \frac{\theta}{2}}\,.
 \label{170637_9Mar21} 
\end{equation}

Let us now evaluate $\mathcal{N}_1$ by substituting $\tilde{G}^{(0)}$ of Eq.~\eqref{170637_9Mar21} into Eq.~\eqref{164812_9Mar21}. Noting here that the integration contour $C'$ reduces to an integration with respect to $\theta$ on a circle far from the core of the vortex, we have
\begin{align}
 \mathcal{N}_1  &= \frac{2\left( 5\cdot 4 \right) ~\epsilon_{ijk}}{5 ! (2\pi)^3} \int \df^3 p \int \df p_4
\int _0 ^{2\pi}\df \theta \nn
&\quad  \times 
\mr{Tr}_D
\Bigg[
\left(\tilde{G}^{(0)} \frac{\partial}{\partial {p_i}}  \left(\tilde{G}^{(0)} \right)^{-1} \right)
\left(\tilde{G}^{(0)} \frac{\partial}{\partial {p_j}}  \left(\tilde{G}^{(0)}\right)^{-1} \right) \nn
&\qquad\qquad \times
\left(\tilde{G}^{(0)} \frac{\partial}{\partial {p_k}}  \left(\tilde{G}^{(0)} \right)^{-1} \right) 
\left(\tilde{G}^{(0)} \frac{\partial}{\partial {p_4}} \left(\tilde{G}^{(0)} \right)^{-1} \right)
\left(\tilde{G}^{(0)} \frac{\partial}{\partial {\theta}} \left(\tilde{G}^{(0)} \right)^{-1}\right)
\Bigg] \, .
\end{align}
Here, $\epsilon_{ijk}$ is the anti-symmetric tensor with $\epsilon_{123}=1$. From the explicit evaluations as
\begin{subequations}
\begin{align}
\tilde{G}^{(0)} \frac{\partial}{\partial {\theta}} \left(\tilde{G}^{(0)} \right)^{-1} &=
e^{-i n\gamma_5 \frac{\theta}{2}} \left(\gamma^\mu p_\mu - m_D \right)^{-1}  (-i nm_D \gamma^5) e^{i n\gamma_5 \frac{\theta}{2}} \, ,
\\
\tilde{G}^{(0)} \frac{\partial}{\partial p_4} \left(\tilde{G}^{(0)} \right)^{-1} &=
e^{-i n\gamma_5 \frac{\theta}{2}} \left(\gamma^\mu p_\mu - m_D \right)^{-1}  (-i)\gamma^0 e^{i n\gamma_5 \frac{\theta}{2}} \, ,\\
\tilde{G}^{(0)} \frac{\partial}{\partial p_i} \left(\tilde{G}^{(0)} \right)^{-1} &=
e^{-i n\gamma_5 \frac{\theta}{2}} \left(\gamma^\mu p_\mu - m_D \right)^{-1}  \gamma^i e^{i n\gamma_5 \frac{\theta}{2}} \, ,
\end{align}
\end{subequations}
we obtain
\begin{align}
 \mathcal{N}_1 & = \frac{-2\epsilon_{ijk}}{6(2\pi)^2i} \int \df^3 p \int \df p_4
~\mr{Tr}_D
\left[
\frac{1}{\slashed{p}-m_D} \gamma^i
\frac{1}{\slashed{p}-m_D} \gamma^j
\frac{1}{\slashed{p}-m_D} \gamma^k
\frac{1}{\slashed{p}-m_D} i\gamma^0
\frac{1}{\slashed{p}-m_D} (-i nm_D \gamma^5)
\right] \nn
&= \frac{-2n\epsilon_{ijk} m_D^2}{6(2\pi)^2} \mr{Tr}_D
\left[
i\gamma^5
 \gamma^i
 \gamma^j
 \gamma^k
  \gamma^0
\right]
 \int \df^3 p \int_{-\infty}^\infty \df p_4
\frac{1}{(-p^2+m_D^2)^3} \,.
\end{align}
Here, we can easily perform the $p_4$- and $\bm{p}$-integrations such that
\begin{equation}
 \int \df^3 p\int_{-\infty}^\infty \df p_4 \frac{1}{(p_4^2 +\bm{p}^2+m_D^2)^3} 
 = \frac{3\pi}{8}\times (4\pi)\int_0^\infty \df |\bm{p}|  \frac{\bm{p}^2}{(\bm{p}^2+m_D^2)^{5/2}}
=\frac{(2\pi)^2}{8m_D^2} \,.
\end{equation}
Together with $\mr{Tr}_D
\left[i\gamma^5 \gamma^i\gamma^j\gamma^k\gamma^0\right]=4\epsilon^{ijk}$ and  $ \epsilon_{ijk}\epsilon^{ijk}=6$, the topological invariant ends up as
\begin{align}
 \mathcal{N}_1 
& = \frac{-2nm_D^2}{6(2\pi)^2} \times (4  \epsilon_{ijk}\epsilon^{ijk} )
\times \frac{(2\pi)^2}{8m_D^2}
 = -n\, .
\end{align}
Therefore, it turns out that this calculation actually reproduces the well-known result given by Ref.~\cite{Jackiw:1981ee} without solving the Dirac equation.\footnote{
We comment on the relation between the above result and the bulk topological invariant introduced in Appendix~\ref{sec:momentum_space_topology_3d}.
The former can be rewritten as $ \mathcal{N}_1=|\mathcal{N}_{CT}|$,
from which it is clear that the number of the zeromodes $\mathcal{N}_1$ is related with 
the bulk topological invariant $\mathcal{N}_{CT}$ (Eq.~\eqref{eq:N_CT}) protected by parity symmetry.
Therefore, the vortex zeromode is understood via the bulk-edge correspondence,
namely, the existence of the edge modes are ensured by the non-vanishing bulk topological invariant $\mathcal{N}_{CT}$.
Note that, $ \mathcal{N}_1$ itself does not depend on the sign of the Dirac mass $m_D$.
}

From this argument, one may predict the number of the zeromodes arising from the electron sector 
in the presence of the $Z$-string background~\eqref{eq: $Z$-string ansatz}.
Because it obtains the Dirac mass from the Yukawa coupling with the SM Higgs field in Eq.~\eqref{LSM},
the Green's function is the same as Eq.~\eqref{eq: Green's function of NO string} with $m_D$ replaced by $h' v_\Phi/\sqrt{2}$.
Thus, the number of the zeromodes is $\mathcal{N}_1=-n$ for the $Z$-string of the vorticity $n$.

\subsection{Application 1: Type-II seesaw model}
\label{sec:top-typeII}
In this subsection, we evaluate the topological invariant \eqref{164812_9Mar21} for the type-II seesaw model. In Subsection~\ref{subsubsection: Calculation of topological invariant}, it turns out that 
the topological invariant for the neutrino sector entails a non-vanishing $\mathcal N_1$. This implies that the net topological invariant \eqref{eq: total vorticity} vanishes due to the cancellation of the contributions from the electron and neutrino sectors. Therefore, the $Z$-string is not stabilized. To see this fact from another viewpoint, in Subsection~\ref{sec:sol_typeII}, we explicitly construct non-trivial solutions to the equation of motion for the left-handed neutrino field and show that indeed there exist such zeromode solutions corresponding to two Majorana-Weyl fermions.

\subsubsection{Calculation of topological invariant}
\label{subsubsection: Calculation of topological invariant}

As a more non-trivial example, we consider the type-II seesaw model for neutrinos. Its Lagrangian is given by
\begin{align}
\mathcal L_\text{Type-II} &= \mathcal L_{\rm SM}\big|_{\lambda\to0} 
+ \Tr [(D^\mu \Delta)(D_\mu \Delta)] + V(\Phi, \Delta) - \left( \frac{1}{2}(y_{M})_{ij} (\ell_i^c)^\dagger i\tau^2\Delta \ell_j + \text{h.c.} \right)
\,,
\label{eq: Type-II seesaw model}
\end{align}
where $\mathcal L_{\rm SM}\big|_{\lambda\to0}$ is the SM Lagrangian given in Eq.~\eqref{LSM} without the potential term proportional to $\lambda$ and Eq.~\eqref{R neutrino sector}, respectively, and $V(\Phi,\Delta)$ denotes the potential for the $SU(2)$-doublet Higgs field $\Phi$ and the $SU(2)$-triplet Higgs field $\Delta=\Delta^a \tau^a/2$ with $\tau^a$ being the Pauli matrices ($a=1,2,3$). The triplet-Higgs field has $U(1)_Y$ hypercharge $Y=2$, so that the covariant derivative acting on it is given by $D_\mu \Delta=\p_\mu\Delta -ig[\frac{\tau^a}{2}, \Delta]W^a_\mu - ig'B_\mu$ and the potential for the Higgs fields can contain $V(\Phi,\Delta)\ni \Phi^T i\tau^2 \Delta^\dagger \Phi + {\rm h.c.}$. In this model, the left-handed lepton doublet fields $\ell_i$ ($i=1,2,3$) couple to the triplet Higgs field as Majorana Yukawa interactions $(y_M)_{ij}$, where $\ell^c$ is the charge conjugation of the lepton doublet fields,
$\ell^c = i \sigma^2 \ell^\ast$. 
Hereafter, for simplicity, we consider the one-flavor case ($i=1$) in which Yukawa coupling matrices are reduced to single coupling constants.

Spontaneous symmetry breaking is realized when the $SU(2)$ double- and triplet-Higgs fields take the vacuum expectation values such that
\begin{align}
&\langle \Phi \rangle = \frac{1}{\sqrt{2}}\pmat{0 \\ v_\Phi}\,,&
&\langle \Delta \rangle = \frac{1}{\sqrt{2}}\pmat{0 && 0 \\ v_\Delta && 0}\,.
\label{eq: VEV of Higgs fields}
\end{align}
On this vacuum, the Majorana Yukawa terms in Eq.~\eqref{eq: Type-II seesaw model} turn to the Majorana-mass terms for the left-handed neutrino
\begin{align}
\frac{M_L}{2}(\nu_L^c)^\dagger \nu_L + \frac{M_L^*}{2}\nu_L^\dagger \nu_L^c\,,
\end{align}
with $M_L=y_M v_\Delta/\sqrt{2}$.

Now, the $Z$-string ansatz is given by the Higgs double and $Z$ fields in \eqref{eq: $Z$-string ansatz}
and the new triplet field $\Delta$,
\begin{align}
\Delta  = \frac{v_\Delta}{\sqrt{2}}\pmat{0 && 0 \\  f_\Delta(r) e^{2in\theta}&& 0}\,,
\label{eq:string-typeII}
\end{align}
where the functions $f(r)$ and $\zeta(r)$ in \eqref{eq: $Z$-string ansatz} and $f_\Delta (r)$ in \eqref{eq:string-typeII} should be determined by solving the EOMs from the Lagrangian \eqref{eq: Type-II seesaw model}.
Note that the vorticity of the triplet Higgs field is given by $2n$ which reflects its hypercharge $Y=2$. This can be understood also from the term $\Phi^T i\tau^2 \Delta^\dagger \Phi + {\rm h.c.}$ in $V(\Phi,\Delta)$. Therefore, in the $Z$-string background
with vorticity $n$, the Majorana mass equips vorticity $2n$. 

Thanks to the topological nature, we can turn off the gauge and Yukawa interactions. 
Thus, the non-interacting Lagrangian for the left-handed neutrino sector reads
\begin{align}
 \mathcal{L}_{\rm lh\nu} & = \nu_L ^\dagger \bar{\sigma}^\mu i \partial_\mu \nu_L - \frac{M_L}{2}\left(e^{i (2n)\theta} (\nu_L^c)^\dagger \nu_L + e^{-i (2n)\theta}\nu_L^\dagger \nu_L^c \right)  \nn
&= \frac{1}{2}\overline N_L \gamma^\mu i \partial_\mu N_L - \frac{M_L}{2}\overline{N}_L e^{-i (2n)\theta\gamma^5} N_L \, ,
\end{align}
where we have introduced the Nambu-Gorkov (four-component) spinors
\begin{equation}
 N_L=\begin{pmatrix}
    \nu_L \\ \nu_L^c
   \end{pmatrix} \, ,
\quad \overline N_L \equiv -N_L^T i\gamma^2 \gamma^0 \, .
\label{eq: Nambu-Gorkov spinors}
\end{equation}

Now, one can see that the Lagrangian looks similar to that of the Dirac fermion coupled with the ANO vortex.
By doing the same calculation given in the last subsection,
we obtain
\begin{equation}
 \label{011956_10Mar21}
\mathcal{N}_1 = \frac{1}{2}\times (2n) = n\,.
\end{equation}
Note that in general the net degrees of freedom of the Nambu-Gorkov spinor are a half of a Dirac spinor, 
so that $\mathcal{N}_1$ gets a factor $1/2$ compared to those of the Dirac fermion.
This means that a minimal unit localized on the vortex is a half of a single Weyl fermion
i.e., a Majorana-Weyl fermion in two dimensions.
However, in the type-II seesaw model~\eqref{eq: Type-II seesaw model}, 
because the left-handed Majorana mass is originated from the VEV of the triplet field $\Delta$ whose $U(1)_Y$ hypercharge is $2$, the vorticity of the left-handed Majorana mass has to be $2n$ under a $Z$-string background of the vorticity $n$. 
Therefore, the topological invariant~\eqref{011956_10Mar21} yields $1$ for $n=1$. 
It implies that inside the $Z$-string, there are always two Majorana-Weyl fermions which form a single Weyl fermion.

On the other hand, the electron sector also gives a single Weyl fermion with the opposite chirality. Together with the neutrino sector, the net Dirac index \eqref{eq: total vorticity} is found to be zero as $\mathcal{N}_1=n_+-n_-=0$ with $n_+=1$ (neutrino) and $n_-=1$ (electron) for the $Z$-string in the type-II seesaw model.
The Dirac index in the quark sector also vanishes due to the same reason as the cancellation between zeromodes in up and down sectors.
Consequently, the $Z$-string is not stabilized by fermion zeromodes in the type-II seesaw model.\footnote{
Here, we comment on the ``bulk-edge correspondence'' in the current model setup. As discussed in Appendix~\ref{sec:momentum_space_topology_3d}, one can define the bulk topological invariant $\mathcal N_K$ where $K$ is an appropriate operation acing on spinor space associated to a discrete symmetry the system has. See Eq.~\eqref{eq:N_K}. The type-II seesaw model is invariant under time-reversal ($T$) symmetry (or $CP$ symmetry). Then, the topological quantity \eqref{011956_10Mar21} is related with $N_T$ such that  
\begin{align}
 \mathcal{N}_1 = | \mathcal{N}_T|\times  2n\,.
\end{align}
The bulk topological invariant $\mathcal N_T$ is explicitly calculated in Eq.~\eqref{eq:NT-typeII}. This relation ensures the existence of zeromodes on the $Z$-string as edge modes in terms of the bulk topology. 
}

\subsubsection{Explicit solution}
\label{sec:sol_typeII}
This emergence of the Majorana-Weyl fermion on the $Z$-string can be verified explicitly by solving the Dirac equation.
When we ignore the gauge coupling and the Yukawa coupling with the Higgs doublet, the Dirac equation for $\nu_L$ in the presence of the $Z$-string \eqref{eq: $Z$-string ansatz} and \eqref{eq:string-typeII} reads 
\begin{equation}
 i \bar{\sigma}^\mu \partial _\mu \nu_L = M_L f_\Delta(r) e^{-2ni\theta} \nu_L^c \, .
 \label{eq: Dirac equation for neutrino}
\end{equation}
In the cylindrical coordinate $(t,r,\theta,z)$, where the vortex core is placed at $r=\sqrt{x^2+y^2}=0$,
Eq.~\eqref{eq: Dirac equation for neutrino} is written as 
\begin{equation}
\left[ 
i \begin{pmatrix} 0 & e^{-i\theta} \\ e^{i\theta} & 0\end{pmatrix} \partial _r
+ i \begin{pmatrix} 0 & -i e^{-i\theta} \\ i e^{i\theta} & 0\end{pmatrix} \frac{1}{r}\partial _\theta 
+ i ( -\partial _0 + \sigma^3 \partial_z)
\right]\nu_L 
= -M_L f_\Delta(r) e^{-2ni\theta} \begin{pmatrix} 0 & 1 \\ -1 & 0\end{pmatrix}\nu_L^\ast \, .\label{eq:polarcoord-Dirac-eq}
\end{equation}
From this, it is obvious that the upper (lower) component corresponds to the left-mover (right-mover) mode traveling on the vortex
since the $z$-$t$ direction in the Dirac equation is easily solved by multiplying a factor $e^{i \omega (z\pm t)}$ with an arbitrary constant $\omega$.

\begin{figure}
    \centering   \includegraphics[width=0.6\columnwidth]{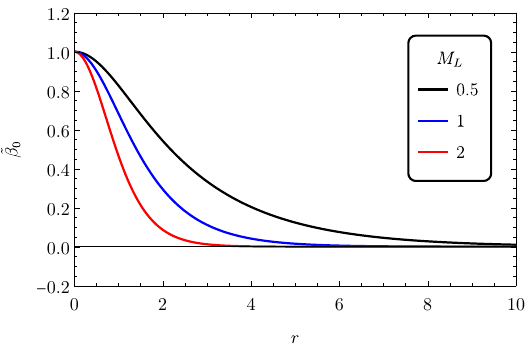}
    \caption{
 The solutions $\tilde\beta_0(r)$ to Eq.~\eqref{214719_9Mar21} with $\tilde\beta_0(0)=1$ in cases of $M_L=0.5$ (black), $1$ (blue) and $2$ (red).
 For $f_\Delta(r)$, we use the numerical solution of $f (r)$ with $\beta \, (= M_H^2 / M_Z^2) = 1$ and the winding number being $1$.
    }
    \label{fig:typeII_1}
\end{figure}

To solve Eq.~\eqref{eq:polarcoord-Dirac-eq},
let us first take $n=1/2$.
This value is not allowed in the $Z$-string configuration since the Higgs doublet becomes a non-single-valued field $\propto e^{i\theta/2}$.
However, it is noteworthy to consider that case to demonstrate 
that the Dirac equation provides a single Majorana-Weyl zeromode.
Focusing on $\omega=0$, the mode expansion for $\nu_L$,
\begin{equation}
\nu_L =  \sum_{m = -\infty}^\infty e^{i m\theta}
\begin{pmatrix}
 \alpha_m(r) \\[1ex]
 \beta_m (r) 
\end{pmatrix}\, ,\label{eq:mode-expand-nuL}
\end{equation}
reduces the Dirac equation into two sets of equations 
\begin{subequations}
\begin{align}
    \alpha_m' - \frac{m}{r} \alpha_m &= - iM_L f_\Delta(r) \alpha_{- 2 - m}^\ast \,,\\
    \alpha_{- 2 - m}' + \frac{2 + m}{r}  \alpha_{- 2 - m} &= - i M_L f_\Delta(r) \alpha_{m}^\ast\,, \\
    \beta_m' + \frac{m}{r} \beta_m &= i M_L f_\Delta(r)  \beta_{- m}^\ast\,, \\
    \beta_{- m}' - \frac{m}{r} \beta_{- m} &= i M_L f_\Delta(r)  \beta_{m}^\ast \, .
\end{align}
\end{subequations}
Let us here concentrate on the case of $m=0$. 
Thus, we arrive at
\begin{subequations}
\begin{align}
 \alpha_0' &= - i M_L f_\Delta(r) \alpha_{-2}^\ast\,,
  \label{211551_9Mar21} \\
  \alpha_{-2}' + \frac{2}{r} \alpha_{-2} &= - i M_L f_\Delta(r) \alpha_0^\ast\,,
  \label{230316_9Mar21}\\
 \beta_0' &= i M_L f_\Delta(r) \beta_0^\ast \, .
 \label{230348_9Mar21}
\end{align}
\end{subequations}
The first two equations \eqref{211551_9Mar21} and \eqref{230316_9Mar21} have no regular solution except for $\alpha_0(r)= \alpha_2(r)=0$ (see Ref.~\cite{Starkman:2000bq}).
On the other hand, the third one \eqref{230348_9Mar21} can entail a non-trivial solution. To see this, we make a redefinition $\beta_0 \equiv e^{-i\frac{\pi}{4}} \tilde{\beta}_0$ in order to remove the coefficient ``$i$'' in equation and decompose $\tilde{\beta}_0$ into the real and imaginary parts,
$\tilde{\beta}_0= \tilde{\beta}_{0,R} + i \,\tilde{\beta}_{0,I}$,
for which Eq.~\eqref{230348_9Mar21} reads
\begin{subequations}
\begin{align}
\tilde{\beta}_{0,R}' &= - M_L f_\Delta(r) \tilde{\beta}_{0,R}\,,
\label{214719_9Mar21} \\
\tilde{\beta}_{0,I}' &= M_L f_\Delta(r) \tilde{\beta}_{0,I}\,. \label{eq:alpha0_imag}
\end{align}
\end{subequations}
Since Eq.~\eqref{eq:alpha0_imag} does not have any normalizable solution except for $\tilde{\beta}_{0,I}=0$,
these solutions are found to be
\begin{equation}
 \tilde{\beta}_0 (r) = \tilde{\beta}_{0,R} \propto \exp\left[-M_L \int _0 ^r \df r'~ f_\Delta(r')\right]\,.
\end{equation}
This solution is displayed in Fig.~\ref{fig:typeII_1}.
Therefore, we obtain a zeromode solution as
\begin{equation}
 \nu_L \propto \begin{pmatrix}
	  0 \\ \tilde{\beta}_0 (r) 
	 \end{pmatrix}\,, \quad \left(\tilde{\beta}_0^\ast=\tilde{\beta}_0\right)
  \label{eq:minimal-sol-typeII-nuL}
\end{equation}
up to an overall constant defined by the normalization condition.
Note that this only has the lower component,
which is a right-mover fermion traveling on the string.
In addition, this solution has only one real degree of freedom,
and hence behaves as a Majorana particle in two dimensions.
Therefore, we have obtained one Majorana-Weyl zeromode.

\begin{figure}
    \centering
    \includegraphics[width=0.6\columnwidth]{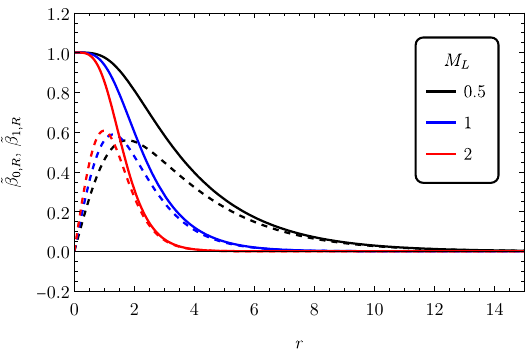}
    \caption{
 The solutions $\tilde\beta_{0,R}(r)$ (solid) and $\tilde\beta_{1,R}(r)$ (dashed) to Eqs.~\eqref{eq:typeII_alpha2_real} with $\tilde\beta_0(0)=1$ in cases of $M_L=0.5$ (black), $1$ (blue) and $2$ (red).
 For $f_\Delta(r)$, we use the numerical solution of $f (r)$ with $\beta \, (= M_H^2 / M_Z^2) = 1$ and the winding number being $2$.
    }
    \label{fig:typeII_2}
\end{figure}

We now come back to the case of the $Z$-string with integer $n$.
For the single $Z$-string, $n=1$, 
the equation of motion \eqref{eq:polarcoord-Dirac-eq} yields 
\begin{subequations}
\begin{align}
    \alpha_m' - \frac{m}{r} \alpha_m &= - i M_L f_\Delta  \alpha_{- 3 - m}^\ast \,,\\
    \alpha_{- 3 - m}' + \frac{3+m}{r} \alpha_{3 - m} &= - i M_L f_\Delta \alpha_m^\ast\,,\\ 
   \beta_m' + \frac{m}{r} \beta_m &= i M_L f_\Delta \beta_{- 1 - m}^\ast\,,
\label{eq:typeII_alpha1}\\
   \beta_{- 1 - m}' + \frac{- 1 - m}{r} \beta_{- 1 - m} &= i M_L f_\Delta  \beta_m^\ast \, .
\label{eq:typeII_alpha2}
\end{align}
\end{subequations}
Although there is no regular solution for $\alpha_m$ again,
$\beta_m$ has a regular solution for $\beta_0$ and $\beta_1$.
To see this, we again redefine as $\beta_0 \equiv e^{-i\frac{\pi}{4}} \tilde{\beta}_0$ and $\beta_1 \equiv e^{-i\frac{\pi}{4}} \tilde{\beta}_1$,
and decompose them into real and imaginary parts.
Then we obtain two pairs of coupled real equations,
\begin{subequations}
\begin{equation}
    \begin{cases} \displaystyle
  \tilde{\beta}_{0,R}' = - M_L f_\Delta  \tilde{\beta}_{1,R} \,, \\[2ex]
      \displaystyle
  \tilde{\beta}_{1,R}' - \frac{1}{r} \tilde{\beta}_{1,R} = - M_L f_\Delta  \tilde{\beta}_{0,R} \, , \\
  \end{cases} \label{eq:typeII_alpha2_real}
  \end{equation}
\begin{equation}
 \begin{cases}  \displaystyle
  \tilde{\beta}_{0,I}' = M_L f_\Delta  \tilde{\beta}_{1,I}\,,\\[2ex]
 \displaystyle
  \tilde{\beta}_{1,I}' - \frac{1}{r} \tilde{\beta}_{1,I} =  M_L f_\Delta  \tilde{\beta}_{0,I} \, . \label{eq:typeII_alpha2_imag}
\end{cases}
\end{equation}
\end{subequations}
The numerical solutions to Eqs.~\eqref{eq:typeII_alpha2_real} are shown in Fig.~\ref{fig:typeII_2}.
Obviously, Eqs.~\eqref{eq:typeII_alpha2_imag} have almost the same solutions as Eqs.~\eqref{eq:typeII_alpha2_real},
but the relative sign between $\tilde{\beta}_{0,I}$ and $\tilde{\beta}_{1,I}$ is opposite to that of $\tilde{\beta}_{0,R}$ and $\tilde{\beta}_{1,R}$.
Therefore, 
denoting particular \textit{real} solutions of Eqs.~\eqref{eq:typeII_alpha2_real} by $\tilde{\beta}_{0,R}=\chi_0(r)$ and $\tilde{\beta}_{1,R}=\chi_1(r)$,
we can express the general zeromode solutions as
\begin{equation}
 \begin{pmatrix}  \displaystyle
    \tilde{\beta}_0 \\ \tilde{\beta}_1
\end{pmatrix}
=
C_1
\begin{pmatrix} \displaystyle
\chi_0(r) \\ \chi_1(r)
\end{pmatrix}
+ i \, C_2
\begin{pmatrix} \displaystyle
\chi_0(r) \\ -\chi_1(r)
\end{pmatrix}
\end{equation}
with $C_1,C_2$ being \textit{real} arbitrary constants.
This is a superposition of two independent real solutions, which are two Majorana-Weyl fermions with the same chirality (right-mover) in two dimensions and equivalent to a single Weyl fermion.
Therefore, we have the index of the zeromodes $\mathcal{N}_1=1$,
which is consistent with the result in Eq.~\eqref{011956_10Mar21}.

\subsection{Application 2: Type-I seesaw model}
\label{sec:top-typeI}
In the same manner as Subsection~\ref{sec:top-typeII}, we calculate the topological invariant for the neutrino sector in the case of the type-I seesaw model. The neutrino sector again produces a non-zero topological invariant. As is the case in the type-II seesaw model, the net topological invariant \eqref{eq: total vorticity} in the type-I seesaw model becomes zero. Hence, the present of fermion zeromodes cannot stabilize the $Z$-string. The fact that there are non-trivial zeromode solutions in the neutrino sector has been discussed in Ref.~\cite{Starkman:2000bq}. In Subsection~\ref{sec: Explicit solution (review)}, we review the zeromode solutions.

\subsubsection{Calculation of topological invariant}
The next example is the type-I seesaw model whose Lagrangian is given by
\begin{align}
\mathcal L_\text{Type-I}= \mathcal L_{\rm SM} + \mathcal L_{\nu_R}\,.
\end{align}
See Eqs.~\eqref{LSM} and \eqref{R neutrino sector} for their explicit forms.
Let us consider the case in which the Dirac mass is originated from the Dirac-Yukawa interaction $\sim (hv_\Phi/\sqrt{2} ) f(r)e^{in\theta}$ with the winding number $n$.
In the bulk far from the center of the vortex,
the Lagrangian for the neutrino sector becomes
\begin{equation}
 \mathcal{L}_\nu= \nu_L ^\dagger \bar{\sigma}^\mu i \partial_\mu \nu_L + \nu_R ^\dagger \sigma^\mu i \partial_\mu \nu_R \, 
 - \left(\frac{M_R}{2} (\nu_R^c)^\dagger \nu_R + \mr{h.c.} \right) 
 - m_D \left(\nu_R^\dagger e^{in\theta} \nu_L +  \nu_L^\dagger e^{-in\theta} \nu_R\right) \, ,
\end{equation}
where $m_D=hv_\Phi/\sqrt{2}$. It is assumed here that the right-handed Majorana mass has no vorticity.
When we write the mass term in the mass-matrix form as
\begin{equation}
\frac{1}{2}
 \begin{pmatrix}
 \nu_L^\dagger & (\nu_R^c) ^\dagger
 \end{pmatrix}
 \begin{pmatrix}
 0 & m_D e^{-in\theta} \\ m_D e^{-in\theta} & M_R
 \end{pmatrix}
 \begin{pmatrix}
 \nu_L^c \\ \nu_R  
 \end{pmatrix} + \mr{h.c.} \, ,
\end{equation}
it is clear that we can vary the Majorana mass $M_R$ from finite values to zero without closing the massgap (in other words, keeping the determinant of the mass matrix as non-zero).
Therefore, the topological invariant $\mathcal{N}_1$ in this case 
is proved to be equal (with the opposite sign) to that in the Dirac case in the presence of the ANO vortex,
i.e.,
\begin{equation}
 \mathcal{N}_1 = n \,. \label{eq:N1_typeI}
\end{equation}
This is consistent with the result in Ref.~\cite{Starkman:2000bq}, in which a Weyl zeromode is obtained for $n=1$.

\subsubsection{Explicit solution (review)}
\label{sec: Explicit solution (review)}
Let us obtain the zeromode explicitly, which have been argued in Refs.~\cite{Starkman:2000bq}. 
The equations of motion for the neutrino fields read from Eqs.~\eqref{LSM} and \eqref{R neutrino sector} as
\begin{align}
i\bar \sigma^\mu D_\mu \nu_L =&  h\phi_d^\ast \nu_R\,, 
\\[1ex]
i\sigma^\mu \partial_\mu \nu_R =& h\phi_d \nu_L + M_R (\nu_R)^c \,,
\end{align}
where we have supposed the $Z$-string ansatz~\eqref{eq: $Z$-string ansatz}, i.e. the covariant derivative acting on the left-handed neutrino field is given by $D_\mu \nu_L=(\partial_\mu - iq Z_\mu) \nu_L$ with Eq.~\eqref{eq: $Z$-string ansatz}.

The Dirac equations are explicitly rewritten as
\begin{align}
\left[ 
i \sigma^r \partial _r
+ i \sigma^\theta \frac{1}{r}\left(\partial _\theta + in \zeta(r)\right)
+ i ( -\partial _0 + \sigma^3 \partial_z)
\right]\nu_L 
&= -m_D f(r) e^{-in\theta} \nu_R \, , \label{eq: EoM of nuL} \\
\left[ 
i \sigma^r \partial _r
+ i \sigma^\theta \frac{1}{r} \partial _\theta 
+ i ( \partial _0 + \sigma^3 \partial_z)
\right]\nu_R 
&= m_D f(r) e^{in\theta} \nu_L - M_R \,i\sigma^2 \nu_R ^\ast \, . \label{eq: EoM of nuR}
\end{align}
From this, it is obvious that the lower (upper) component of $\nu_R$ corresponds to the left-mover (right-mover) mode traveling on the vortex, which are opposite to those of $\nu_L$.
The $z$-$t$ direction is easily solved by multiplying a factor $e^{i \omega (z\pm t)}$ with an arbitrary constant $\omega$.
Again, we focus on the zero-energy solution $\omega=0$.

In order to solve Eqs.~\eqref{eq: EoM of nuL} and \eqref{eq: EoM of nuR}, we decompose them into the differential equations for each component of the spinor fields,
\begin{align}
&\nu_L =\pmat{ \alpha \\ \beta }\,,&
&\nu_R =\pmat{ \gamma \\ \delta}\,.
\label{eq: left and right hand neutrinos}
\end{align}
Plugging Eq.~\eqref{eq: left and right hand neutrinos} into Eqs.~\eqref{eq: EoM of nuL} and \eqref{eq: EoM of nuR}, we obtain the equations of motion for each component of the neutrino fields. Those of the left-handed neutrino field read
\begin{align}
ie^{i\theta}\left[ \p_r - \frac{1}{r} \Big(-i\p_\theta + n \zeta(r)\Big) \right]\alpha &= \left( - m_D f(r)  e^{-in\theta} \right)\delta\,,\\[1ex]
ie^{-i\theta}\left[ \p_r + \frac{1}{r} \Big(-i\p_\theta + n \zeta(r) \Big) \right]\beta &=  \left( - m_D f(r)e^{-in\theta} \right)\gamma \,,
\end{align}
while for the right-handed neutrino field, we have
\begin{align}
-ie^{i\theta}\left[ \p_r + \frac{i}{r}\p_\theta  \right]\gamma + M_R \gamma^* &= \left( - m_D  f(r)  e^{in\theta}\right)\beta\,,\\
-ie^{-i\theta}\left[ \p_r - \frac{i}{r}\p_\theta  \right]\delta - M_R \delta^*&=  \left( - m_D  f(r) e^{in\theta}\right)\alpha\,.
\end{align}
To solve these coupled differential equations, we perform the mode expansions for each component as
\begin{subequations}
\label{fullansatz}
\begin{align} 
\alpha(t,r,\theta,z)  =& \sum_{m=-\infty}^{\infty}  \alpha_m(r) e^{i m\theta}\,, \\
\beta(t,r,\theta,z)  =& -i \sum_{m=-\infty}^{\infty}  \beta_m(r) e^{i m\theta}\,, \\
\gamma(t,r,\theta,z)  =& \sum_{m=-\infty}^{\infty}   \gamma_m(r) e^{i m\theta}\,, \\
\delta(t,r,\theta,z)  =& -i \sum_{m=-\infty}^{\infty}   \delta_m(r) e^{i m\theta}\,,
\end{align}
\end{subequations}
and then obtain the equations of motion for $\alpha_m(r)$, $\beta_m(r)$, $\gamma_m(r)$ and $\delta_m(r)$.

Hence, we look for solutions to the following recursive equations with the winding number $n=1$:
\begin{subequations}
\begin{align}
\alpha'_m - \frac{(m+\zeta)}{r}\alpha_m  =& m_D f \delta_{m+2} \,,\\[1ex]
\beta'_m + \frac{(m+\zeta)}{r}\beta_m  = & -m_D f \gamma_{m}\label{EoM for beta}\,, \\[1ex]
\gamma'_m - \frac{m}{r} \gamma_m + iM_R \gamma^\ast_{-1-m} =& - m_D f \beta_{m}\label{EoM for gamma}\,, \\[1ex]
\delta'_m + \frac{m}{r} \delta_m + iM_R \delta^\ast_{1-m} =&  m_D f \alpha_{m-2} \,.
\label{EoM for delta}
\end{align}
\end{subequations}
In the $M_R=0$ limit, by replacing $m$ with $m+2$ in Eq.~\eqref{EoM for delta}, these differential equations become closed forms for $\alpha_m$, $\beta_m$, $\gamma_m$ and $\delta_{m+2}$ which are the standard equations for zeromodes of the Dirac particle.

We now focus on Eqs.~\eqref{EoM for beta} and \eqref{EoM for gamma}.
The replacement of $m$ with $-1-m$ gives
\begin{align}
\beta'_{-1-m} + \frac{-1-m+\zeta}{r}\beta_{-1-m}  = & - m_D f \gamma_{-1-m}\,,\label{eq: EoM for beta -1-m} \\[1ex]
\gamma'_{-1-m} - \frac{-1-m}{r} \gamma_{-1-m} + iM_R \gamma^\ast_{m} =& - m_D f \beta_{-1-m}\label{eq: EoM for gamma -1-m} \,.
\end{align}
Thus, the four equations, Eqs.~\eqref{EoM for beta}, \eqref{EoM for gamma}, \eqref{eq: EoM for beta -1-m}, and \eqref{eq: EoM for gamma -1-m}, are closed.
We eliminate $\gamma_m$ and $\gamma_{-1-m}$ from Eqs.~\eqref{EoM for gamma} and \eqref{eq: EoM for gamma -1-m} by using Eqs.~\eqref{EoM for beta} and \eqref{eq: EoM for beta -1-m}
and then have
\begin{align}
&\frac{\df}{\df r}
\left[
\frac{1}{m_D f}
\left(
\tilde{\beta}'_{m} + \frac{m+\zeta}{r}\tilde{\beta}_{m} 
\right) 
\right]
\nonumber \\
&-
\frac{m}{r}
\frac{1}{m_D f}
\left(
\tilde{\beta}'_{m} + \frac{m+\zeta}{r}\tilde{\beta}_{m}
\right) 
-
M_R \left(\tilde{\beta}'^\ast_{-1-m} + \frac{-1-m+\zeta}{r}\tilde{\beta}_{-1-m}^\ast \right)
= m_D f \tilde{\beta}_{m}\,,
\label{eq:betam}
\\
&\frac{\df}{\df r}
\left[
\frac{1}{m_D f}
\left(
\tilde{\beta}'_{-1-m} + \frac{-1-m+\zeta}{r}\tilde{\beta}_{-1-m}
\right)
\right]
\nonumber \\
&+
\frac{1+m}{r}
\frac{1}{m_D f}
\left(
\tilde{\beta}'_{-1-m} + \frac{-1-m+\zeta}{r}\tilde{\beta}_{-1-m}
\right) 
-
M_R
\left(
\tilde{\beta}'^\ast_{m} + \frac{m+\zeta}{r}\tilde{\beta}^\ast_{m}
\right)
= m_D f \tilde{\beta}_{-1-m}\,,
\label{eq:betaMmM1}
\end{align}
where we have again redefined as $\beta_m \equiv e^{-i\frac{\pi}{4}} \tilde{\beta}_m$ and $\gamma_m \equiv e^{-i\frac{\pi}{4}} \tilde{\gamma}_m$ for convenience.
Note that all coefficients are real, so that we can separate the equations into real and imaginary parts.

\begin{figure}
    \centering
    \includegraphics[width=0.6\columnwidth]{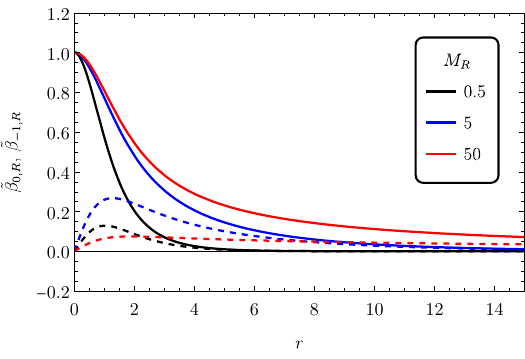}
    \caption{
    Plots for the zeromode solutions $\tilde \beta_{0,R}$ and $\tilde \beta_{-1,R}$ for $M_R = 0.5$ (black), $5$ (blue), and $50$ (red) as functions of $r$ in the energy unit $M_Z/\sqrt{2}=1$.
    The Dirac mass is taken to be $m_D = 1$ in this unit.
    The solid and dashed lines stand for $\tilde\beta_0(r)$ and $\tilde\beta_{-1}(r)$, respectively.
    For $f(r)$, we take $\beta \, (= M_H^2 / M_Z^2) = 1$.
    }
    \label{fig:typeI}
\end{figure}

Let us here solve the real parts $\tilde{\beta}_{0,R}$ and $\tilde{\beta}_{-1,R}$ of the above coupled two equations numerically. 
For this, it is actually sufficient to deal with the equations for $m=0$.
We fix the background bosonic fields ($Z_\mu$ and $\phi_d$) as the $Z$-string configuration. For various $M_R$, we find that the equations always have solutions for the real parts $\tilde \beta_{0,R} (r)$ and $\tilde \beta_{-1,R} (r)$. Fig.~\ref{fig:typeI} displays these zeromode solutions for $M_R=0.5$, $5$, $50$ (in the energy unit $M_Z/\sqrt{2}=1$).
One can see from Fig.~\ref{fig:typeI} particularly that for large $M_R$, the solution becomes broader, namely slowly converges for $r\to \infty$.
This is because each of the solutions asymptotically becomes a linear combination of two exponential tails with typical mass scales $M_R$ and $m_D^2 / M_R$, as stated in Ref.~\cite{Starkman:2000bq}. For large $M_R$, the former tail ($\sim e^{-M_Rr}$) suddenly decay, while the latter tail ($\sim e^{-\left( m_D^2 / M_R\right)r}$) survives even with large $r$, which provides the broad profile of the solutions.
As $M_R \to \infty$, the solutions eventually become non-normalizable.
This is consistent with the decoupling limit, that is, in the limit of $M_R\to \infty$ the model reduces to the SM without the right-handed neutrino, and hence the neutrino sector in the SM does not allow normalizable zeromodes on the electroweak string.

For the imaginary parts $\tilde \beta_{0,I}$ and $\tilde \beta_{-1,I}$, we get almost the same solutions as the real ones except for that their relative sign is opposite to that of $\tilde \beta_{0,R}$ and $\tilde \beta_{-1,R}$.
Therefore, repeating the same argument presented in Sec.~\ref{sec:sol_typeII},
it is found that we have obtained the two independent real solutions for $(\tilde \beta_0, \tilde \beta_{-1})$,
corresponding to a right-mover Weyl fermion in two dimensions.
Note that $(\tilde \gamma_0,\tilde\gamma_{-1})$ are not independent from $(\tilde \beta_0, \tilde \beta_{-1})$
since they are determined by Eqs.~\eqref{EoM for beta} and \eqref{eq: EoM for beta -1-m}.
In addition, the solutions for $\alpha$ and $\delta$ are not found except for the trivial ones, $\alpha_m=\delta_m=0$.
Therefore, we conclude that the index $\mathcal{N}_1$ in the background of the single $Z$-string is given as $\mathcal{N}_1=1$,
which is consistent with Eq.~\eqref{eq:N1_typeI} and Ref.~\cite{Starkman:2000bq}.

\subsection{Application 3: Type-I seesaw model with one \texorpdfstring{$\nu_L$}{} and two \texorpdfstring{$\nu_R$}{}'s}
Now, we can consider a \textit{modified} type-I seesaw model
in which there are one left-handed and two right-handed neutrinos.\footnote{If one considers a gauge $\mathbb{Z}_4$ symmetry under which the SM fermions and the right-handed neutrinos are charged, 
this setup suffers from the Dai-Freed anomaly~\cite{Dai:1994kq,Garcia-Etxebarria:2018ajm}.
In such a case, there should be another spectator fermion to cancel this anomaly.
We do not consider the gauge $\mathbb{Z}_4$ symmetry in this paper.
}
The neutrino sector is extended to be the following Lagrangian
\begin{align}
 \mathcal{L}_\text{neutrino}&= \nu_L ^\dagger \bar\sigma^\mu i \partial_\mu \nu_L + \sum_{j=1,2} \nu_{R,j} ^\dagger \,  \sigma^\mu i \partial_\mu \, \nu_{R,j} \nn
&\qquad
-  \sum_{j,k} \left(\frac{M_{jk}}{2} (\nu_{R,j}^c)^\dagger \nu_{R,k} + \mr{h.c.} \right) 
 -  \sum_{j}\left(m_j \nu_{R,j}^\dagger \nu_L + \mr{h.c.} \right) \, ,
\end{align}
where the two right-handed neutrinos are labeled by the index $j$ or $k$. The Majorana and Dirac masses are described by a $2\times 2$ matrix $M_{jk}$ and a $1\times 2$ matrix (vector) $m_j$.
The mass term is expressed by the following $3 \times 3$ mass-matrix form:
\begin{equation}
 \begin{pmatrix}
 \nu_L^\dagger & (\nu_{R,1}^c) ^\dagger & (\nu_{R,2}^c) ^\dagger
 \end{pmatrix}
 \begin{pmatrix}
 0 & m_1 & m_2 \\ 
m_1 & M_{11} & M_{12} \\
m_2 & M_{21} & M_{22} 
 \end{pmatrix}
 \begin{pmatrix}
 \nu_L^c \\ \nu_{R,1} \\ \nu_{R,2}
 \end{pmatrix} + \mr{h.c.} \, .
\end{equation}

Let us consider the case in which the two Dirac masses $m_1$ and $m_2$ have a vorticity $n$, 
i.e., $m_j\, e^{in\theta}$ for $j=1,2$. 
We calculate the topological invariant describing the number of the zeromodes, $\mathcal{N}_1$.
Without loss of generality, we can diagonalize the Majorana mass matrix $M_{jk}$, leading to the mass matrix
\begin{equation}
\mathcal M\equiv
 \begin{pmatrix}
 0 & m_1 e^{-in\theta}& m_2e^{-in\theta} \\ 
m_1 e^{-in\theta} & M_1 & 0 \\
m_2 e^{-in\theta} & 0 & M_2 
 \end{pmatrix} \, ,
\end{equation}
and can take $M_1,M_2 \geq0$.
Because $\det \mathcal M \propto -M_1 m_2^2 - M_2 m_1^2$,
we put $m_1=M_2=0$ without closing the massgap.
Note that we cannot take $M_1=0$ at the same time.

The Dirac equations are given as
\begin{subequations}
\begin{align}
& i \bar \sigma^\mu \partial _\mu \nu_L =  m_2 e^{-in\theta} \nu_{R,2}\,,\\ 
& i \sigma^\mu \partial _\mu \nu_{R,1} = M_1 \nu_{R,1}^c \,,\\
& i \sigma^\mu \partial _\mu \nu_{R,2} =  m_2 e^{in\theta} \nu_L \,.
\end{align}
\end{subequations}
From this, it is obvious that $\nu_{R,1}$ is decoupled from the system
and $\nu_{R,2}$ and $\nu_L$ form one species of Dirac fermion.
Therefore, the topological invariant $\mathcal{N}_1$ is equal to that in the Dirac case in the presence of the ANO vortex (with the opposite sign),
i.e.,
\begin{equation}
 \mathcal{N}_1 = n \,.
\end{equation}
To obtain the zeromode explicitly, we can use the same argument for the Dirac fermion.
See Subsection~\ref{sec: Dirac fermion on Abrikosov-Nielsen-Olesen vortex}.

\subsection{Application 4: Hybrid of type-I and type-II seesaw models}
Let us consider the hybrid of the type-I and type-II seesaw.
We concentrate on one generation of the SM fermions and one right-handed neutrino.
The Lagrangian is given by a sum of $\mathcal{L}_\text{Type-II}$ \eqref{eq: Type-II seesaw model} and 
$\mathcal{L}_{\nu_R}$ \eqref{R neutrino sector}.
The $Z$-string background is the same as that in the type-II seesaw model,
\eqref{eq: $Z$-string ansatz} and \eqref{eq:string-typeII}.
Thanks to the topological nature, 
we turn off the gauge and Yukawa interactions again,
and thus the neutrino sector is described by the following Lagrangian
\begin{align}
 \mathcal{L}_\text{Type-I+Type-II} 
& = \nu_L ^\dagger \bar\sigma^\mu i \partial_\mu \nu_L + \nu_R ^\dagger  \sigma^\mu i \partial_\mu \nu_R 
- \frac{M_L}{2}\left(e^{i (2n)\theta} (\nu_L^c)^\dagger \nu_L + e^{-i (2n)\theta}\nu_L^\dagger \nu_L^c \right)  \n \\
& \hspace{3em} - \left(\frac{M_R}{2} (\nu_R^c)^\dagger \nu_R + \mr{h.c.} \right) 
 - m_D \left(\nu_R^\dagger e^{in\theta} \nu_L +  \nu_L^\dagger e^{-in\theta} \nu_R\right)  \, ,
\end{align}
which leads to the mass matrix of the neutrinos,
\begin{equation}
\frac{1}{2}
 \begin{pmatrix}
 \nu_L^\dagger && (\nu_R^c) ^\dagger
 \end{pmatrix}
 \begin{pmatrix}
 M_L e^{-2ni\theta} && m_D e^{-in\theta} \\[1ex] m_D e^{-in\theta} && M_R
 \end{pmatrix}
 \begin{pmatrix}
 \nu_L^c \\[1ex] \nu_R  
 \end{pmatrix} + \mr{h.c.} \, .
\end{equation}
Note that the left-handed Majorana and the Dirac masses has the winding phase, while the right-handed Majorana mass does not.

In contrast to the case given in Subsection~\ref{sec:top-typeI},
now we can turn off neither of $M_L$ nor $M_R$,
and the system can be classified into two phases corresponding to
\begin{equation}
 M_L M_R - m_D^2 \gtrless 0 \, .
\end{equation}
These are topologically distinct phases 
since in order to move from one phase into the other phase, 
one needs to get across the phase boundary on which the mass gap is closed.

Let us look at what value of $\mathcal{N}_1$ each topological phase has.
For $ M_L M_R - m_D^2 > 0$, 
one can take $m_D=0$ and $M_L,M_R>0$ as a representative point.
In this case, the right-handed neutrino $\nu_R$ is completely decoupled from the SM sector,
and the setup reduces to the case of Type-II seesaw model (discussed in Subsection~\ref{sec:top-typeII}),
resulting in $\mathcal{N}_1=n$.
On the other hand, for $ M_L M_R - m_D^2 < 0$, 
one can take $M_L=M_R=0$ without loss of generality,
and hence we have the same setup as the usual Dirac fermion in the background of the ANO string, resulting in $\mathcal{N}_1=n$.
Therefore, both cases have the same value $\mathcal{N}_1=n$.

\subsection{Comment on the stability of \texorpdfstring{$Z$}{}-string}
So far, we have shown that
in any seesaw model of the Type-I, Type-II, their hybrid, and the Type-I with different flavors,
we have the same number of neutrino zeromodes on the $Z$-string background, $\mathcal{N}_1=n$ with $n$ the vorticity.
Accompanying with the electron sector, we have the total number of zeromodes in the lepton sector (for one generation) as
\begin{align}
    \mathcal{N}_1 &= n~\text{(neutrino)} - n~\text{(electron)} =0\,,
\end{align}
which means that the $1+1$-dimensional effective theory consisting of the zeromodes on the string is not a chiral theory but a vector-like one with the same degrees of freedom of $n$ (two-component) Dirac fermions.
As stated above, the zero Dirac index on a string indicates the instability of the string itself 
because a pair of the fermion zeromodes with opposite chiralities forms a massive mode lowering the energy
under the perturbative deformation of the string.
Therefore, what we have found indicates that the $Z$-string cannot be stabilized by any extension of the neutrino sector in the SM.

On the other hand, in Ref.~\cite{Starkman:2001tc},
it is shown that in the same setup as the last subsection (hybrid of type-I and -II),
the number of the zeromodes from the neutrino sector can vanish,
which means that the Dirac index from the neutrino and electron sectors becomes non-zero in total.
It is further argued that this non-zero index can stabilize the $Z$-string.
At a first look, our result is inconsistent with the result in Ref.~\cite{Starkman:2001tc}.
We here give some comments on the apparent contradiction.

Firstly, one should note that this model admits another string solution 
when one imposes a global $U(1)_L$ ($L$ denotes the lepton number) symmetry associated with the phase rotation of the SM leptons and the triplet $\Delta$.
This symmetry prohibits the cubic term $\Phi^T i\tau^2 \Delta^\dagger \Phi$ in the potential.
The $U(1)_L$ symmetry is spontaneously broken by the VEV $v_\Delta$,
which leads to a global string solution described by the following ansatz
\begin{align}
\Phi = \frac{v_\Phi}{\sqrt{2}}\pmat{0 \\  1}\,, &
& \Delta  = \frac{v_\Delta}{\sqrt{2}}\pmat{0 & 0 \\  f_\Delta(r) e^{i\ell\theta}& 0}\,, &
&Z_\theta = 0 \, ,
\label{eq:lepton-string}
\end{align}
with the vorticity $\ell\in \mathbb{Z}$ of the global string.
Since the gauge coupling does not matter, the Dirac equation for $\nu_L$ in the background of this string is given as
\begin{equation}
 i \bar\sigma^\mu \partial _\mu \nu_L = M_L f_\Delta(r) e^{-i\ell\theta} \nu_L^c \, ,
\end{equation}
which is equivalent to the Dirac equation for the $Z$-string~\eqref{eq: Dirac equation for neutrino}
with $\ell$ replaced by $2n$.
Note that, in this case, the minimal vorticity is $|\ell|=1$ instead of $|\ell|=2$
because the Higgs doublet is not charged under the $U(1)_L$ symmetry
and is a single-value field for arbitrary $\ell$.
Taking the minimal value $\ell=1$,
which is equivalent to substituting $n=1/2$ into Eq.~\eqref{eq: Dirac equation for neutrino},
then we again get the unique solution \eqref{eq:minimal-sol-typeII-nuL}.
Therefore the single global $U(1)_L$ string provides a single Majorana-Weyl fermion, or equivalently, a half of a Weyl fermion.

Next, let us point out that, in the study in Ref.~\cite{Starkman:2001tc}, 
they consider a background described as follows
\begin{align}
&\Phi \sim \frac{1}{\sqrt{2}}
\begin{pmatrix}
 0 \\ v_\Phi e^{i\theta}
\end{pmatrix}\,,&
&\Delta \sim \frac{1}{\sqrt{2}}
\begin{pmatrix}
 0 & 0 \\ v_\Delta& 0
\end{pmatrix} \, ,
& Z_\theta \sim \frac{1}{q} \, ,
\end{align}
which is different from our $Z$-string ansatz~\eqref{eq: $Z$-string ansatz} and \eqref{eq:string-typeII}
in the sense that the triplet Higgs $\Delta$ does not contain the winding phase.
Instead, this is nothing but a superposition of the single $Z$-string and two global $U(1)_L$ strings with the phase of $\Delta$ anti-rotated,
i.e., the vorticity for $Z$-string is $1$, while that for the global $U(1)_L$ string is $\ell=-2$.
In fact, in Ref.~\cite{Starkman:2001tc} it is mentioned that this configuration gives a logarithmic divergent tension from the covariant derivative $D_\mu \Delta$ to be $4\times 2\pi v_\Delta^2\log \Lambda$ with $\Lambda$ being the infrared cutoff,
which is a feature of the global string.

In this case, thanks to the topological nature, 
the Dirac index on the string is given by the sum of contributions from the single $Z$-string and the two global $U(1)_L$ strings,\footnote{This is understood by considering a case in which these strings are separated far enough so that each of them is regarded as a single string.}
\begin{align}
    \mathcal{N}_1 &= \underbrace{1}_\text{$\nu$ zeromodes on $Z$-string} - \underbrace{1}_\text{$e$ zeromodes on $Z$-string} - \underbrace{2 \times \frac{1}{2}}_\text{$\nu$ zeromodes from $U(1)_L$ string} \nn
    & = -1\,.
\end{align}
This is nothing but the value found in Ref.~\cite{Starkman:2001tc}.
This states that this superposed configuration is in the same topological sector with the two global $U(1)_L$ strings without $Z$ flux,
which implies that it evolves into the latter to reduce the energy.
Therefore, it is not shown in Ref.~\cite{Starkman:2001tc} that the $Z$-string itself is stabilized by the fermion zeromodes.
In addition, because the global $U(1)_L$ symmetry cannot be an exact symmetry in the phenomenological viewpoint,
the global $U(1)_L$ string is not stable in the phenomenologically viable setup.
To conclude, we cannot expect any stable string object in the SM with the extensions in the neutrino sector.


\section{Discussion and Conclusions}
\label{sec:discussion_conclusions}

In this paper, we have applied the notion of topological invariants to various extensions of the SM, focusing on the emergence of zeromodes in $Z$-strings.
In Sec.~\ref{sec:zeromode_review} we first have reviewed the electroweak string (or $Z$-string) solution in the SM.
In Sec.~\ref{sec:momentum_space_topology_1d} we have introduced a topological invariant defined in the momentum space and real space, that counts the number of fermion zeromodes along one-dimensional topological defects.
While these solutions can be obtained explicitly by numerical methods, the language of topological invariant makes it easier to understand why and how these zeromodes appear.
We have applied this formalism to several models with extensions in the neutrino sectors
and have found that most seesaw models (type-I, type-II, their hybrid, and modified type-I),
the number of the zeromode from the neutrino sector on the $Z$-string is always $n$ with $n$ the vorticity of the $Z$-string.
We have also critically examined claims in the literature~\cite{Starkman:2001tc} from the viewpoint of these topological invariants.
While it was claimed in the literature that the fermion zeromodes help to stabilize the $Z$-string, 
we pointed out that the stabilization is not the case: the topological object they consider is in fact the superposition of $Z$-string and a global string corresponding to $U(1)_L$, and the former is not stabilized by the zeromodes. 
To conclude, there remains a huge room to apply topological arguments to various models in high-energy physics and consider cosmological implications.

As the simplest extensions of the SM, 
the two Higgs doublet models (2HDMs) have been studied extensively. According to the fermion sectors,  
these models are further classified into several types.
Similarly to the SM, 
the 2HDMs also admit $Z$-strings 
\cite{La:1993je,Earnshaw:1993yu,Bimonte:1994qh},
which are non-topological and are unstable. 
The fermion zeromodes in 2HDMs were not studied yet, and they may or may not appear in various ways depending on the types of 2HDMs.  
A possibility of the stabilization by fermion zeromodes in this case is one of future directions.
In addition, the 2HDMs admit 
topological $Z$-strings 
which are attached by domain walls 
\cite{Dvali:1993sg,Eto:2018hhg,Eto:2018tnk}
as axion string, and can support 
topological Nambu monopoles \cite{Eto:2019hhf,Eto:2020hjb,Eto:2020opf}.
The stabilization of a non-topological $Z$-string 
by splitting into a pair of topological $Z$-strings was also proposed in Ref.~\cite{Eto:2021dca}. 
The fermion zeromodes may enhance the stability region in such a mechanism.
In a phenomenological viewpoint in particle physics, the number of fermion generations might be explained in analogy to the number of zeromodes in topological insulators~\cite{Kaplan:2011vz}.

In condensed matter physics, it is known that the number of zeromodes (edge modes) is strongly related with the topology of the bulk momentum space without defects, as the so-called bulk-edge correspondence.
Thus, it can be expected that the topological invariant we have studied in this work would be understood in the context of the bulk-edge correspondence.
This might help us to classify particle physics models in terms of the Altland-Zirnbauer classification~\cite{altland1997nonstandard} for symmetry-protected topological phases in condensed matter.

\section*{Acknowledgements}

The authors are grateful to Ken Shiozaki for helpful comments.
This work is supported in part by 
the WPI program ``Sustainability with Knotted Chiral Meta Matter (SKCM$^2$)'' at Hiroshima University (M.\,E. and M.\,N.), 
JSPS KAKENHI Grant Numbers 
JP22H01221 (M.\,E. and M.\,N.),
JP21J01117 (Y.\,H.),
JP23K17687 (R.\,J.),
JP23K19048 (R.\,J.), and
the National Science Foundation of China (NSFC) under Grant No.~12205116 (M.\,Y.). 
This work is also supported by 
the Deutsche Forschungsgemeinschaft under Germany's Excellence Strategy - EXC 2121 Quantum Universe - 390833306
and the Seeds Funding of Jilin University.

\appendix
\section{Notation}
\label{sec:notation}
\subsection{Representation of Dirac matrices}
We clarify the representation of the Dirac matrices. In the chiral representation, the Dirac matrices in Cartesian coordinates are given by
\begin{equation}
 \gamma^\mu = \begin{pmatrix}
	       \bm{0} & \sigma^\mu \\ \bar \sigma^\mu  & \bm{0}
	      \end{pmatrix}\,,
\hspace{2em}
\sigma^\mu = (\bm{1},\sigma^i)\,, 
\hspace{2em}
\bar \sigma^\mu = (\bm{1},-\sigma^i)\,,
\label{eq:gamma_mat in chiral}
\end{equation}
for which the chirality matrix becomes
\begin{equation}
 \gamma^5 =i\gamma^0\gamma^1\gamma^2\gamma^3
 =\begin{pmatrix}
	       -\bm{1} & \bm{0} \\ \bm{0} & ~\bm{1}~
\end{pmatrix}\,,
\label{eq:gamma5_mat in chiral}
\end{equation}
and the charge conjugation matrix $C$ reads 
\begin{equation}
C = -i \gamma^0 \gamma^2 = \pmat{ 0&~1~&0&0 \cr -1&0&0&0 \cr
0&0&0&-1 \cr 0&0&~1~&0} \, .
\end{equation}
Since the $Z$-string configuration is obtained in  cylindrical coordinates $(r,\theta, z)$, it is convenient to work in the same coordinates for the neutrino field configurations. 
More specifically, we intend to write
\begin{align}
\gamma^0\p_0 + \gamma^1\p_1+ \gamma^2\p_2 + \gamma^3\p_3 =
\gamma^0\p_0 +\gamma^r \p_r +   \gamma^\theta \frac{1}{r}\p_\theta  +   \gamma^3 \p_3 \,.
\end{align}
Here, a representation of the  Dirac matrices in cylindrical coordinates is found to be
\begin{align}
\label{Gamma}
\gamma^r=
\pmat{  0 & 0 & 0 & e^{-i\theta} \cr 
0 & 0 & e^{i\theta} & 0 \cr
0 & -e^{-i\theta} & 0 & 0 \cr
-e^{i\theta} & 0 & 0  & 0 }\,,\qquad
\gamma^\theta=\pmat{  
0 & 0 & 0 & -ie^{-i\theta} \cr 
0 & 0 & ie^{i\theta} & 0 \cr
0 &  ie^{-i\theta} & 0 & 0\cr
-ie^{i\theta} & 0 & 0 & 0 }\,,
\end{align}
while $\gamma^0$ and $\gamma^3$ are the same as in Eq.~\eqref{eq:gamma_mat in chiral}.

For the two-by-two matrices $\sigma^\mu$,
we have
\begin{align}
\sigma^\mu \partial_\mu=
\sigma^0 \partial_0 + \sigma^1 \partial_1 + \sigma^2 \partial_2 + \sigma^3 \partial_3
 = \sigma^0 \partial_0 + \sigma^r \partial_r + \sigma^\theta \frac{1}{r}\partial_\theta + \sigma^3 \partial_3
\end{align}
with
\begin{align}
 \sigma^r = 
\begin{pmatrix}
 0 & e^{-i\theta} \\
e^{i\theta} & 0
\end{pmatrix} \, ,
\quad
 \sigma^\theta = 
\begin{pmatrix}
 0 & -ie^{-i\theta} \\
ie^{i\theta} & 0
\end{pmatrix} \, .
\end{align}

\section{Bulk momentum space topology in neutrino sector}
\label{sec:momentum_space_topology_3d}

In this appendix, we consider the neutrino models in light of topology. We first introduce another topological invariant defined in the momentum space of the neutrino fields by closely following Refs.~\cite{Groves:1999ks,Volovik:2016mre,Zubkov:2016llc}.

\subsection{Topological invariant for symmetry-protected topological phases}
\label{app:subsec:Topological invariant}
Let us start by reviewing general prescriptions to define topological invariants characterizing topological phases protected by symmetry of gapped systems.
Since topological properties are robust under continuous deformations, we switch off interaction in the system, so that we can specify ourselves into one-particle states.
Let $\mathcal{G}$ be a free Green's function with a non-zero mass gap defined on the four-dimensional momentum space $p_\mu$
and $K$ be an operator associated with a discrete symmetry that satisfies $K^2=1$ and acts on the Green's function and (one-particle) Hamiltonian.
Then, one can define the following quantity~\cite{Volovik:2003fe,Volovik:2009jf,Volovik:2010kv,Volovik:2016mre,Zubkov:2016llc}:
\begin{equation}
 \mathcal{N}_K \equiv \frac{\epsilon_{ijk}}{24 \pi^2} 
\mathrm{Tr}\left[\int \df^3 p ~ K (\mathcal{G} \partial_{p_i} \mathcal{G}^{-1})( \mathcal{G} \partial_{p_j} \mathcal{G}^{-1})(\mathcal{G} \partial_{p_k} \mathcal{G}^{-1}) \right] \label{eq:N_K} \, ,
\end{equation}
where the trace is performed over indices of the Dirac spinors and other internal degrees of freedom (if exist).
Let $n$ be the dimension over which the trace is taken (i.e., $\mathcal{G}$ is a $n\times n$ matrix).
In order for this integration to be well-defined,
the integral domain should not contain on-shell pole nor branch cut,
and hence we are led to take $p_0=0$.
In this case, $\mathcal{G}$ is regular everywhere and an element of $GL(n,\mathbb{C})$ (general linear group of degree $n$ over complex numbers).

Let us next see why this quantity~\eqref{eq:N_K} plays a role of the topological invariant.
When $\mathcal{G}$ does not commute nor anti-commute with $K$,
this quantity does not have any topological meaning in general.
On the other hand, when $\mathcal{G}$ satisfies
\begin{equation}
 \mathcal{G} K = K \mathcal{G} \quad \text{or}  \quad \mathcal{G} K = -K \mathcal{G}\,,
\end{equation}
one can evaluate the integral \eqref{eq:N_K} within each $K=\pm 1$ subspace,
\begin{equation}
\left. \mathcal{N}_K \right|_{K=\pm1} = \pm \frac{\epsilon_{ijk}}{24 \pi^2} 
\underset{K=\pm 1}{\mathrm{Tr}}\left[\int \df^3 p ~ (\mathcal{G} \partial_{p_i} \mathcal{G}^{-1})( \mathcal{G} \partial_{p_j} \mathcal{G}^{-1})(\mathcal{G} \partial_{p_k} \mathcal{G}^{-1}) \right] \,,
\end{equation}
where $\underset{K=\pm 1}{\mathrm{Tr}}$ indicates the trace over the subspaces with $K=\pm 1$.
Noting that $GL(n,\mathbb{C})$ homotopes to the unitary group $U(n)$,
this integration resembles the Pontryagin index and gives a topological invariant of a map from the three-dimensional torus $T^3$ into $U(n)$.
Since this quantity is topologically quantized,\footnote{
Because the momentum space considered here is not three-sphere $S^3$ but $T^3$, 
this quantity is different from the Pontryagin index to measure $\pi_3(U(n))= \mathbb{Z}$.
As a result, it is not topological in a strict sense but can be continuously changed when one introduces interactions violating spatial translational symmetry~\cite{PhysRevB.91.245148,Ryu:2010zza}.
Nevertheless, we can practically regard it topological 
because the translational symmetry is usually respected in particle physics setups.
} 
it is useful to characterize (symmetry-protected) topological phases.

\subsection{A simple example: Free Dirac fermion}
\label{sec:top-dirac}
As a toy model, let us consider a free Dirac fermion,
whose Lagrangian is given by
\begin{align}
 \mathcal{L} & = \bar \psi i \gamma^\mu \partial_\mu \psi+ m_D \bar \psi \psi \, .
\end{align}
This Lagrangian is invariant under the $CT$ transformation (which is combination of the charge and time-reversal transformations):
\begin{align}
 \psi \xrightarrow[]{CT} -\gamma^1 \gamma^3 (-i) (\bar \psi \gamma^0\gamma^2)^T
 = -i \gamma^1 \gamma^2 \gamma^3 \psi^\ast = i \gamma^5 \gamma^0 \psi^\ast \, .
 \end{align}

Following Subsection~\ref{app:subsec:Topological invariant}, let us evaluate the quantity \eqref{eq:N_K} with $K=i \gamma^5 \gamma^0$,
\begin{equation}
 \mathcal{N}_{CT} \equiv \frac{\epsilon_{ijk}}{24 \pi^2} 
\mathrm{Tr}\left[\int \df^3 p ~ i \gamma^5 \gamma^0 (\mathcal{G} \partial_{p_i} \mathcal{G}^{-1})( \mathcal{G} \partial_{p_j} \mathcal{G}^{-1})(\mathcal{G} \partial_{p_k} \mathcal{G}^{-1}) \right]_{p_0=0} \label{eq:N_CT} \, ,
\end{equation}
with $\mathcal{G}^{-1}\equiv \gamma^\mu p_\mu + m_D $.
If the Lagrangian were not invariant under the $CT$ transformation, this quantity is not topological. Thus, it is meaningful only when the system respects the symmetry.
One can explicitly check now that the quantity \eqref{eq:N_CT} is indeed a topological invariant as follows:
\begin{align}
 \mathcal{N}_{CT} 
&= \frac{\epsilon_{ijk}}{24 \pi^2} 
\mathrm{Tr}\left[\int \df^3 p ~ \frac{i \gamma^5 \gamma^0(\slashed{p}-m_D) \gamma^i (\slashed{p}-m_D) \gamma^j (\slashed{p}-m_D) \gamma^k}
{(p^2 - m_D^2)^3}\right] _{p_0=0} \,,
\end{align}
where the numerator is simplified to be $i \gamma^5 \gamma^0 \gamma^i\gamma^j\gamma^k (-m_D)(\bm{p}^2 + m_D^2)$
with $\bm{p}$ the three-dimensional momentum ($p^2|_{p_0=0} = - \bm{p}^2$). Using a formula $\mathrm{Tr}[ i\gamma^5 \gamma^0 \gamma^i\gamma^j\gamma^k]=4 \epsilon^{ijk} $ and $\epsilon_{ijk}\epsilon^{ijk}=6$, we have
\begin{align}
 \mathcal{N}_{CT}
&= \frac{1}{\pi^2} 
\int \df^3 p ~\frac{m_D}{(\bm{p}^2+m_D^2)^2}
= \frac{m_D}{|m_D|} 
= {\rm sign}(m_D)\,.
\end{align}
Thus, this is a topological invariant, which only depends on the sign of the Dirac mass. Note that, in this free Dirac fermion system, the sign of the Dirac mass can be easily changed by redefinition of the field $\psi$, which means that $\mathcal{N}_{CT}$ itself cannot have any physical meaning.\footnote{
Once higher order terms with respect to momenta are taken into account in the dispersion relation, the sign of the Dirac mass becomes physically meaningful in relation to the sign of the coefficient of such terms, and one may discuss whether the system is in a topologically (non)trivial phase, see e.g., Ref.~\cite{Shen_Shan_Lu}.
In relativistic field theories in continuum space, such higher order terms may naturally come from regularizations~\cite{Aoki:2023lqp}.
}

\subsection{Type II seesaw model}
We here give a similar quantity in the type-II seesaw model with the left-handed neutrino having a Majorana mass.
Let $\nu_L$ be a two-component spinor field describing the left-handed neutrino in the SM.
We consider free neutrinos and ignore all interactions to the SM gauge and Higgs fields. The non-interacting Lagrangian for the type II seesaw model is given by
\begin{align}
 \mathcal{L} & = \nu_L^\dagger i \bar\sigma^\mu \partial_\mu \nu_L + \left( \frac{M_L}{2} \nu_L^\dagger \nu_L^c + (\text{h.c.})\right)
 = \frac{1}{2}\overline N \left(i\gamma^\mu \p_\mu + M_L\right)N\,,
 \label{eq:Lagrangian2-typeII}
\end{align}
where the four-component Nambu-Gorkov spinor (Majorana spinor) $N$ defined in Eq.~\eqref{eq: Nambu-Gorkov spinors} has been used.
Since the system \eqref{eq:Lagrangian2-typeII} is not invariant under the $CT$ (nor $P$) transformation, the topological invariant \eqref{eq:N_CT} in this model trivially vanishes. However, one can define a similar quantity associated with the $CP$ (or $T$) symmetry.

As discussed in Section~\ref{sec:top-typeII}, the Majorana mass $M_L$ originates from a Majorana-Yukawa interaction in which a $SU(2)$-triplet Higgs field $\Delta$ takes a vacuum expectation value \eqref{eq: VEV of Higgs fields}, although we do not take into account its origin since we are interested in the broken phase and ignore all interactions.
In this basis, the time-reversal symmetry acts on $N$ as
\begin{align}
  N = \begin{pmatrix}
    \nu_L \\ \nu_L^c
   \end{pmatrix} &\xrightarrow[]{T}
\begin{pmatrix}
     \sigma^1 \sigma^3\nu_L \\ \sigma^1 \sigma^3 \nu_L^c
   \end{pmatrix}=
\begin{pmatrix}
     -i \sigma^2 \nu_L \\ -i \sigma^2 i\sigma^2 (\nu_L^\dagger)^T
   \end{pmatrix} 
=\begin{pmatrix}
    - (\nu_L^{c\dagger})^T \\ (\nu_L^\dagger)^T 
   \end{pmatrix}
= \gamma^5 \gamma^0 (N^\dagger)^T\,.
\end{align}
Therefore, we can define a topological invariant with $K=i\gamma^5\gamma^0$,
\begin{equation}
 \mathcal{N}_T \equiv \frac{1}{2}\frac{\epsilon_{ijk}}{24 \pi^2} 
\mathrm{Tr}\left[\int \df^3 p ~ i \gamma^5 \gamma^0 (\mathcal{G} \partial_{p_i} \mathcal{G}^{-1})( \mathcal{G} \partial_{p_j} \mathcal{G}^{-1})(\mathcal{G} \partial_{p_k} \mathcal{G}^{-1}) \right]_{p_0=0}\,,
\end{equation}
with $\mathcal{G}^{-1}\equiv \gamma^\mu p_\mu + M_L $.
This is the same form as $\mathcal{N}_{CT}$ defined by Eq.~\eqref{eq:N_CT} except for the overall factor $1/2$,
so that we can calculate this in the same way,
which leads to
\begin{align}
 \mathcal{N}_T 
&= \frac{1}{2}\frac{M_L}{|M_L|} 
= \frac{1}{2}{\rm sign}(M_L)\,. \label{eq:NT-typeII}
\end{align}
The topological invariant depends on the sign of the Majorana mass.

\subsection{Type I seesaw model}
We here consider the type-I seesaw model,
in which $\nu_L$ and $\nu_R$ are two-component spinors describing the left-handed and left-handed neutrinos, respectively.
The non-interacting Lagrangian is given by
\begin{align}
 \mathcal{L} & = \nu_L^\dagger i \bar \sigma^\mu \partial_\mu \nu_L + \nu_R^\dagger i \sigma^\mu \partial_\mu \nu_R  
+ \left( m_D \nu_L^\dagger \nu_R - \frac{M_R}{2}\nu_R^\dagger \nu_R^c + (\text{h.c.}) \right)\,,
\label{eq:Lagrangian-typeI}
\end{align}
where
\begin{equation}
 \nu_R^c \equiv -i\sigma^2 (\nu_R^\dagger)^T\,, \qquad  
 \nu_L^c \equiv i\sigma^2 (\nu_L^\dagger)^T \, .
\end{equation}
Here, we perform a field rotation with angle $\theta = m_D/M_R$ assuming $|m_D|\ll |M_R|$ such that the neutrino fields mix as 
 \begin{equation}
 \nu_L^c{}' \simeq \nu_L^c + \theta \nu_R\,, 
 \qquad 
 \nu_R' \simeq \nu_R - \theta \nu_L^c \, ,
 \end{equation}
for which the Lagrangian is rewritten as
\begin{align}
 \mathcal{L} & =  \nu_L'^\dagger i \bar \sigma^\mu \partial_\mu \nu_L' + \nu_R'^\dagger i  \sigma^\mu \partial_\mu \nu_R'  
+ \left( \frac{m_D^2}{2M_R} \nu_L'^\dagger \nu_L'^c - \frac{M_R}{2} (\nu_R'^c)^\dagger \nu_R' + \text{h.c.} \right) \,.
\end{align}
Note that we have used $\nu_L'^T (\nu_R'^\dagger)^T = -\nu_R'^\dagger \nu_L' $ since they are Grassmann variables.

Now, the signs of the mass terms for $\nu_L'$ and $\nu_R'$ are opposite, so that the total topological invariant vanishes:
\begin{align}
    \mathcal{N}_T=  \frac{1}{2}\,\mathrm{sign} \left(-\frac{m_D^2}{M_R} \right) +  \frac{1}{2}\,\mathrm{sign} (M_R) = 0 \, .
\end{align}

\subsection{Hybrid of type I and II models}
It is straight forward to extend the above result into hybrid models of type I and II,
in which the Lagrangian is given by
\begin{align}
 \mathcal{L} & = \nu_L^\dagger i \bar \sigma^\mu \partial_\mu \nu_L  + \nu_R^\dagger i  \sigma^\mu \partial_\mu \nu_R   + \left( \frac{M_L}{2} \nu_L^\dagger \nu_L^c + \frac{M_R}{2}\nu_R^\dagger \nu_R^c + m_D \nu_L^\dagger \nu_R  + (\mathrm{h.c.}) \right) \, .
\end{align}
By diagonalizing the mass matrix on the field basis $(\nu_L, \nu_R)$,
\begin{equation}
\bm{M}_\nu = 
  \begin{pmatrix}
  M_L & m_D \\ m_D & M_R
 \end{pmatrix}\,,
\end{equation}
one obtains mass eigenvalues,
\begin{align}
 m_1 &= \frac{1}{2}\left[M_L + M_R - \sqrt{(M_L-M_R)^2 + 4 m_D^2}\right]\,,\\
 m_2 &= \frac{1}{2}\left[M_L + M_R + \sqrt{(M_L-M_R)^2 + 4 m_D^2}\right] .
\end{align}
Since the total topological invariant is again given as
\begin{equation}
 \mathcal{N}_{T} 
= \frac{1}{2} \mathrm{sign}(m_1) +\frac{1}{2} \mathrm{sign}(m_2)  \,,
\end{equation}
one can easily see that there are two phases of $\mathcal{N}_{T}$:
\begin{align}
&\begin{cases}
  m_1, m_2 >0 &\Leftrightarrow \, \mathcal{N}_T =1\,, \\  
  m_1, m_2 <0 &\Leftrightarrow \, \mathcal{N}_T = -1 \,,\\  
  m_1< 0 \land m_2 >0  &\Leftrightarrow \, \mathcal{N}_T = 0 \,.
\end{cases}
\end{align}
Note here that $m_1 < m_2$ always holds.

It follows that the transition of the sign of $\mathrm{det}\, \bm{M}_\nu =m_1 m_2$ implies the transition of the phases,
which means that the phases cannot be continuously connected unless at least one mass eigenstate becomes massless (i.e., the mass gap is closed).


\bibliographystyle{JHEP} 
\bibliography{zeromode_refs}
\end{document}